\documentclass[10pt]{article}
\usepackage{graphics}

\newcommand{\eg} {{\it e.g., }}
\newcommand{\eq} { {\rm eq} }
\newcommand{\ie} {{\it i.e., }}
\newcommand{\rmd} { {\rm d} }
\newcommand{\rme} { {\rm e} }

\begin{document}


\title{Kinetics of Surfactant Adsorption:\\
The Free Energy Approach}
\author{
Haim Diamant\thanks
{Present address: The James Franck Institute, The University of
Chicago, 5640 South Ellis Avenue, Chicago, IL 60637, USA.}, 
Gil Ariel 
and David Andelman\thanks{Corresponding
author. Tel:~+972-3-6407239;~ Fax:~+972-3-6422979;
e-mail:~andelman@post.tau.ac.il}\\ School of Physics and Astronomy
\\
     Raymond and Beverly Sackler Faculty of Exact Sciences \\
     Tel Aviv University, Ramat Aviv, 69978 Tel Aviv, Israel}

\date{November 13, 2000}

\maketitle

\begin{abstract}

We review the free energy approach to the kinetics of
surfactant adsorption at fluid/fluid interfaces.
The formalism is applied to several systems.
For non-ionic surfactant solutions the results coincide with
previous models while indicating their limits of validity.
We study the case of surfactant mixtures, focusing on the relation
between the
mixture kinetics and the properties of its individual constituents.
Strong electrostatic interactions in salt-free ionic
surfactant solutions drastically modify
the adsorption kinetics.
In this case the theory accounts for experimental
results which could not be previously understood.
The effect of screening by added salt is studied as well.
Our theoretical predictions are compared with available
experiments.

\end{abstract}
\bigskip

\noindent{\it Keywords}: surface tension, adsorption, surfactants,
interfacial phenomena, kinetics



\section{Introduction}
\setcounter{equation}{0}

The kinetics of surfactant adsorption is a fundamental problem
of interfacial science playing a key role in various processes and
phenomena, such as wetting, foaming and stabilization of liquid
films.
For example, the wetting rate of a substrate by surfactant solutions
was shown to be correlated with the {\em dynamic} surface
tension of the solution, rather than its equilibrium surface tension
\cite{wetting}.
Since the pioneering theoretical work of Ward and Tordai in the
1940s \cite{WT},
the kinetics of surfactant adsorption
has been the object of thorough
experimental and theoretical research \cite{review1}--\cite{review3}.

The problem of adsorption kinetics, being a non-equilibrium one,
poses several theoretical difficulties.
One question concerns the adsorption mechanism at the
interface and its coupling to the kinetics in the bulk solution.
Another important question is related to the definition and
calculation of the time-dependent interfacial tension as
measured in experiments.
Previous theoretical works have addressed these questions
by adding appropriate assumptions to the theory.
Such models can be roughly summarized by the following
scheme:
(i) consider a diffusive transport of surfactant molecules from
a semi-infinite bulk solution (following Ward and Tordai);
(ii) introduce a certain adsorption equation as a boundary condition
at the ~interface;~~
(iii) solve~ for the~ time-dependent~ surface coverage;
(iv) assume that the equilibrium equation of state is valid
also out of equilibrium and calculate the dynamic surface tension.
While certain models take an equilibrium adsorption isotherm as
the interfacial boundary condition \cite{Delahay}--\cite{Joos1},
others use a kinetic equation \cite{Miller}--\cite{Franses}.

The purpose of this article is to review a theoretical approach based on a
free-energy formalism \cite{EPL}--\cite{ourmixture}. The main advantage of
the free-energy approach is that all the equations are derived from a
single functional, thus yielding a more complete and consistent
description of the kinetics in the entire system. The definition and
calculation of the dynamic surface tension results naturally from the
formalism itself, and extension to more complicated interactions can then
follow. In this review we summarize the essence of the free-energy
approach and its application to various systems while skipping most of the
technical calculations. More details can be found in previous publications
\cite{EPL}--\cite{ourmixture}.

The next section presents the general theoretical framework and
basic considerations of our formalism. In the sections that follow
we apply this general scheme to three important examples. First,
the simplest case of a single-component, non-ionic surfactant
solution is considered. We analyze the various stages and
characteristic time scales of the adsorption process. Results of
previous models are recovered as special cases, and their limits
of validity are defined. In the second example the treatment of
the non-ionic case is extended to surfactant mixtures. Experiments
portray a large variety of phenomena specific to mixed systems
\cite{Joos1},\cite{LeVan}--\cite{Siddiqui}. For instance, more
complex dynamic surface tension is observed due to competition
between different species. We focus on the relation between the
adsorption behavior of the mixture and the properties of its
individual constituents. Certain cases are found, where mixing
several surfactant species may lead to significant differences in
the kinetics. The third example concerns ionic surfactant
solutions. In salt-free systems, strong electrostatic interactions
are found to drastically modify the adsorption kinetics and yield
interesting time dependence \cite{Langevin1}--\cite{HuaDESS}.
Using our approach we point out the problems in previous models as
applied to such systems and account for the experimentally
observed behavior. Electrostatic screening caused by added salt is
shown to lead to a kinetic behavior much similar to the non-ionic
case.

Our theoretical predictions are compared to available experiments.
However, specific experimental techniques, as can be found in 
Ref.~\cite{review1}, are not covered. Since a considerable body of
theoretical work is summarized in this review, derivations are not
given in full detail; further details can be found in 
Refs.~\cite{EPL}--\cite{ourmixture}.


\section{Theoretical Framework}
\label{sec_framework}
\setcounter{equation}{0}

This section outlines the general free energy formalism, which is
used extensively in the sections that follow \cite{EPL}.

\begin{figure}[p]
\centerline{ \resizebox{0.57\textwidth}{!}
{\includegraphics{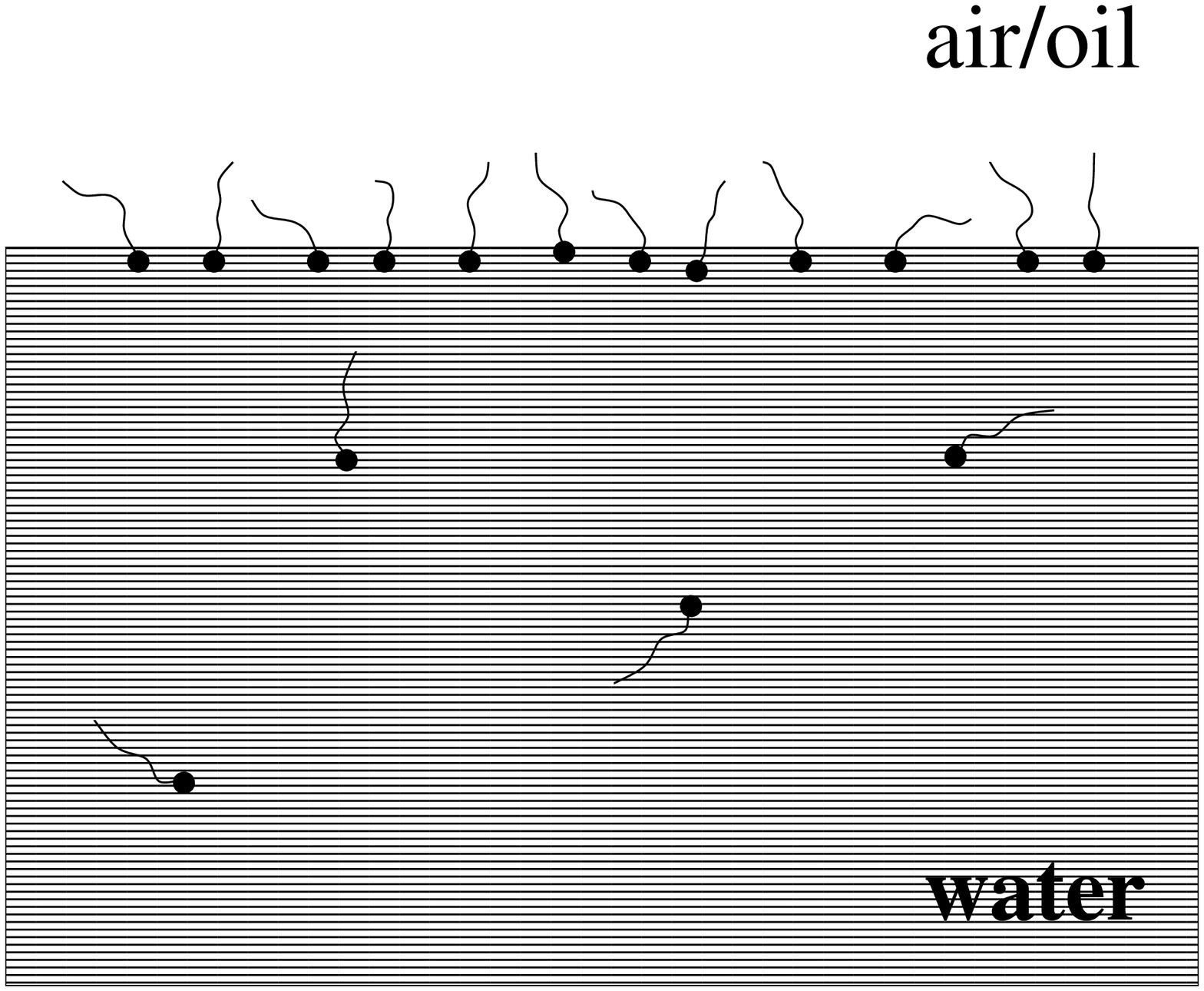}} }
\caption[]{}
\label{fig_system}
\end{figure}

In this review we assume that the aqueous solution has a sharp,
flat interface
with another non-polar fluid phase (an oil or air phase),
as is illustrated in Fig.~\ref{fig_system}.
We are concerned with a surfactant solution below its
critical micelle concentration ({\it cmc}),
\ie containing only monomers.
In such a dilute solution there are two important energy scales:
the thermal energy, $T$ (throughout this review we take the
Boltzmann constant as unity),
and the energy of molecular transfer to the water/oil or water/air
interface, $\alpha$.
In common surfactant systems $\alpha$ is much larger than $T$
(typically in the range 10--20$T$).
As a result, a very compact monolayer is formed at the
interface with an interfacial volume fraction close to unity.
Since the bulk volume fraction in such dilute solutions is very low
(typically $10^{-6}$--$10^{-4}$), the surfactant attains
a step-like profile having a sharp decrease within a molecular
distance from the interface.
It is unjustified in these circumstances to employ a continuum,
gradient-expansion formalism for the entire system, as is done
in many other interfacial problems.
A more appropriate approach is to treat the interface as a distinct
sub-system being in thermal and diffusive contact with the bulk
solution \cite{Tsonop}.
Consequently, the excess free energy of the system is divided
into a bulk contribution and an interfacial one.

We write the excess free energy per unit area
as a functional of the various degrees of freedom,
$\{\phi_i\}$, required to describe the system (\eg the surfactant
volume fraction profile, electrostatic potential, etc.),
\begin{equation}
  \Delta\gamma[\{\phi_i\}] = \int_0^\infty
  \Delta f[\{\phi_i(x,t)\}]{\rm d}x ~+~ f_0 [\{\phi_{i0}(t)\}].
\label{Dg_general}
\end{equation}
In the first term, $\Delta f$ denotes the local excess in free
energy density over the bulk, uniform solution, $x$ being the
distance from the interface and $t$ the time. The second term,
$f_0$, describes the contribution from the interface itself,
$\{\phi_{i0}\}$ being the interfacial values of the various
degrees of freedom. A coupling term is to be included in $f_0$ to
account for the contact between the interface and the bulk. Note
that in order to correctly model the kinetics, the coupling should
be made with the layer in contact, namely the sub-surface layer of
solution ($x\rightarrow 0$), since it is generally {\em not}
in equilibrium with the rest of the bulk during the process.
It has been implicitly assumed in Eq.~\ref{Dg_general} that
lateral inhomogeneities are negligible, \ie the time scale of
lateral kinetics is assumed very short compared to the adsorption
process. This assumption is usually justified for fluid/fluid
interfaces and allows a reduction of the problem to a single
spatial dimension, namely the distance from the interface, $x$
\cite{lateral}.

Apart from $T$ and $\alpha$, another energy parameter
is usually required to quantitatively account for equilibrium
as well as
kinetic experimental measurements \cite{Lin2}.
It is associated with lateral attraction between surfactant
molecules at the interface, which usually cannot be neglected
due to the large interfacial coverage.
Values of surfactant--surfactant interaction parameters may amount to
several $T$, the thermal energy.

Once a free energy functional in the form of
Eq.~\ref{Dg_general} has been formulated,
the equilibrium relations and kinetic equations are easily derived.
Equilibrium relations, such as the equilibrium profile and
adsorption isotherm, are found by setting the variation of the
free energy with respect to the various degrees of freedom to zero,
\begin{equation}
  \frac {\delta\Delta\gamma} {\delta\phi_i(x)} = 0 , \ \ \ \ \
\mbox{equilibrium}.
\label{equilibrium_general}
\end{equation}
The corresponding extremum of the free energy yields the
equilibrium equation of state, relating $\Delta\gamma$ with the
equilibrium values of $\{\phi_i\}$. First-order kinetic equations
can be derived as well from the variation of the free energy.
Since the degrees of freedom relevant to the adsorption problem
are usually conserved quantities (\eg concentration profiles), the
scheme for deriving the kinetic equation for a conserved order
parameter should be employed (see, \eg Ref.~\cite{Langer}),
\begin{equation}
   \frac{\partial\phi_i}{\partial t} = \frac{a_i^2}{T}  \frac{\partial}{\partial x}
   \left[ D_i(\{\phi_i\})
   \phi_i \frac{\partial}{\partial x} \left( \frac{\delta\Delta\gamma}{\delta\phi_i}
   \right) \right],
\label{kinetic_general}
\end{equation}
where $a_i$ is a molecular size and $D_i(\{\phi_i\})$ a diffusion
coefficient. Due to the step-like profile discussed above, a
similar dependence may be assumed for the diffusion coefficient as
well, \ie having a constant value, $D_i$, in the dilute bulk and
possibly a different value, $D_{i0}$, at the interface.
The kinetic equations derived by this procedure do not account
for convective transport. Convection is found to play a
significant role in certain practical cases and experimental
setups \cite{Sutherland}.
More recent experimental techniques,
however, seem to exclude convective effects \cite{review1},
and they will be neglected in the current review.

The distinction between bulk and interface
results in separate (though coupled) kinetic equations for the two
sub-systems.
Correspondingly, two limiting cases naturally arise.
{\em Diffusion-limited adsorption} occurs when the interfacial
kinetics
is much faster than the transport from the bulk.
In this case the interfacial layer may be assumed to maintain
quasi-equilibrium with the sub-surface layer throughout the process.
Consequently, the interfacial kinetic equations are reduced to
equilibrium-like isotherms relating the surface coverage and
sub-surface
volume fraction. They thus serve merely as static boundary
conditions for the kinetic equations in the bulk.
The other limiting case is {\em kinetically limited adsorption},
where the interfacial kinetics becomes the slow, limiting process,
and the bulk may be assumed throughout the process
as maintaining quasi-equilibrium with the changing interface.
Deriving all the kinetic equations from a single functional
allows a more rigorous determination of the conditions
under which such limiting cases hold.
This will be demonstrated in the following sections.

One of the important points in our formalism is that the excess
free energy per unit area (Eq.~\ref{Dg_general}) is {\em
identified} with the measurable reduction in interfacial tension.
Furthermore, assuming that this definition holds at equilibrium as
well as out of equilibrium readily solves the problem of
calculating the dynamic surface tension, which is a fundamental
obstacle in adsorption kinetics. Previous works dealt with this
obstacle by simply assuming that the equilibrium equation of state
can be used for the {\em dynamic} surface tension as well
\cite{Fordham}. Since the equation of state is merely the extremum
of the functional in Eq.~\ref{Dg_general}, using it out of
equilibrium is valid only when the free energy is not too far from
its minimum value. Noting that the dominant term in
Eq.~\ref{Dg_general} is usually the interfacial one, $f_0$,
this requirement is fulfilled when the interface is close to
equilibrium with the sub-surface layer. In other words, the scheme
employed by previous works is valid only for diffusion-limited
adsorption. This observation becomes particularly important in
kinetically limited systems, such as salt-free ionic surfactant
solutions, where our general equation \ref{Dg_general}, rather
than the equation of state, must be used in order to correctly
calculate the dynamic surface tension.

\section{Non-Ionic Surfactants}
\setcounter{equation}{0}

We start with the simplest case of an aqueous solution
containing a single type of non-ionic surfactant \cite{EPL}.
The excess free energy (Eq.~\ref{Dg_general}) can be
rewritten in
this case as a functional of a single degree of freedom --- the
volume fraction profile of the surfactant, $\phi(x,t)$,
\begin{equation}
  \Delta \gamma[\phi] = \int_0^\infty \Delta f[\phi(x,t)] {\rm d}x
  ~+~ f_0[\phi_0(t)],
 \label{Dg}
\end{equation}
where $\phi_0$ is the volume fraction at the interface
(surface coverage).
We assume a contact with a
reservoir, where the surfactant has fixed volume fraction
and chemical potential, $\phi_{\rm b}$ and $\mu_{\rm b}$,
respectively.
Since the solution is dilute, steric and other short-range
interactions between surfactant
molecules are assumed to take place only at the
interfacial layer itself.
Hence, the two contributions to the excess free energy
are written as
\begin{eqnarray}
  \Delta f(\phi)&=&\{ T [ \phi(\ln\phi - 1) -
                     \phi_{\rm b}(\ln\phi_{\rm b} - 1)]  -
   \mu_{\rm b} (\phi - \phi_{\rm b})
    \}/a^3
\label{Df} \\
        f_0(\phi_0)&=&\{ T [ \phi_0\ln\phi_0 +
   (1-\phi_0)\ln(1-\phi_0)]  -
   (\alpha + \mu_1)\phi_0 - (\beta/2)\phi_0^2 \} / a^2,
 \label{f0}
\end{eqnarray}
where $a$ denotes the surfactant molecular size.
The contribution from the bulk contains only the entropy
in the ideal-solution limit and contact with the reservoir.
In the interfacial contribution, however, we have included the
entropy of mixing accounting for the finite molecular size,
a linear term
accounting for the surface activity and contact with the adjacent
solution [$\mu_1\equiv\mu(x\rightarrow 0)$ being the chemical
potential at the sub-surface layer],
and a quadratic term describing short-range lateral attraction
between surfactant molecules at the interface.
Although both $\alpha$ and $\mu_1$ are linearly coupled with the
surface
coverage, their physical origin is quite different --- $\alpha$
is constant in time, characterizing the surface activity of the
specific surfactant,
whereas $\mu_1$ is a time-dependent function participating in the
interfacial kinetics.
By using a quadratic term for lateral attraction
we restrict to the simplest short-range pair interactions.
This simplification is sufficient for describing
the thermodynamics of monolayers in the gaseous and liquid
states.
It is merely a 2nd order term
of an expansion in $\phi_0$, and generalization to more
complicated situations can be made.

\subsection{Equilibrium Relations}

Setting
the variation of the free energy with respect to $\phi(x)$
and $\phi_0$
to zero yields a uniform profile in the bulk,
$\phi(x>0) \equiv \phi_{\rm b}$, and recovers the Frumkin adsorption
isotherm (or the Langmuir one, if $\beta=0$)
at the interface \cite{Adamson},
\begin{equation}
  \phi_0 = \frac {\phi_{\rm b}} {\phi_{\rm b} + {\rm e}^{-(\alpha+\beta\phi_0)/T}}.
 \label{Frumkin}
\end{equation}
Substituting these results in the free energy functional recovers
also the equilibrium equation of state,
\begin{equation}
  \Delta\gamma = [ T\ln(1-\phi_0) + (\beta/2) \phi_0^2 ] / a^2.
 \label{eqstate}
\end{equation}

\subsection{Kinetic Equations}

\begin{figure}[p]
\centerline{ \resizebox{0.57\textwidth}{!}
{\includegraphics{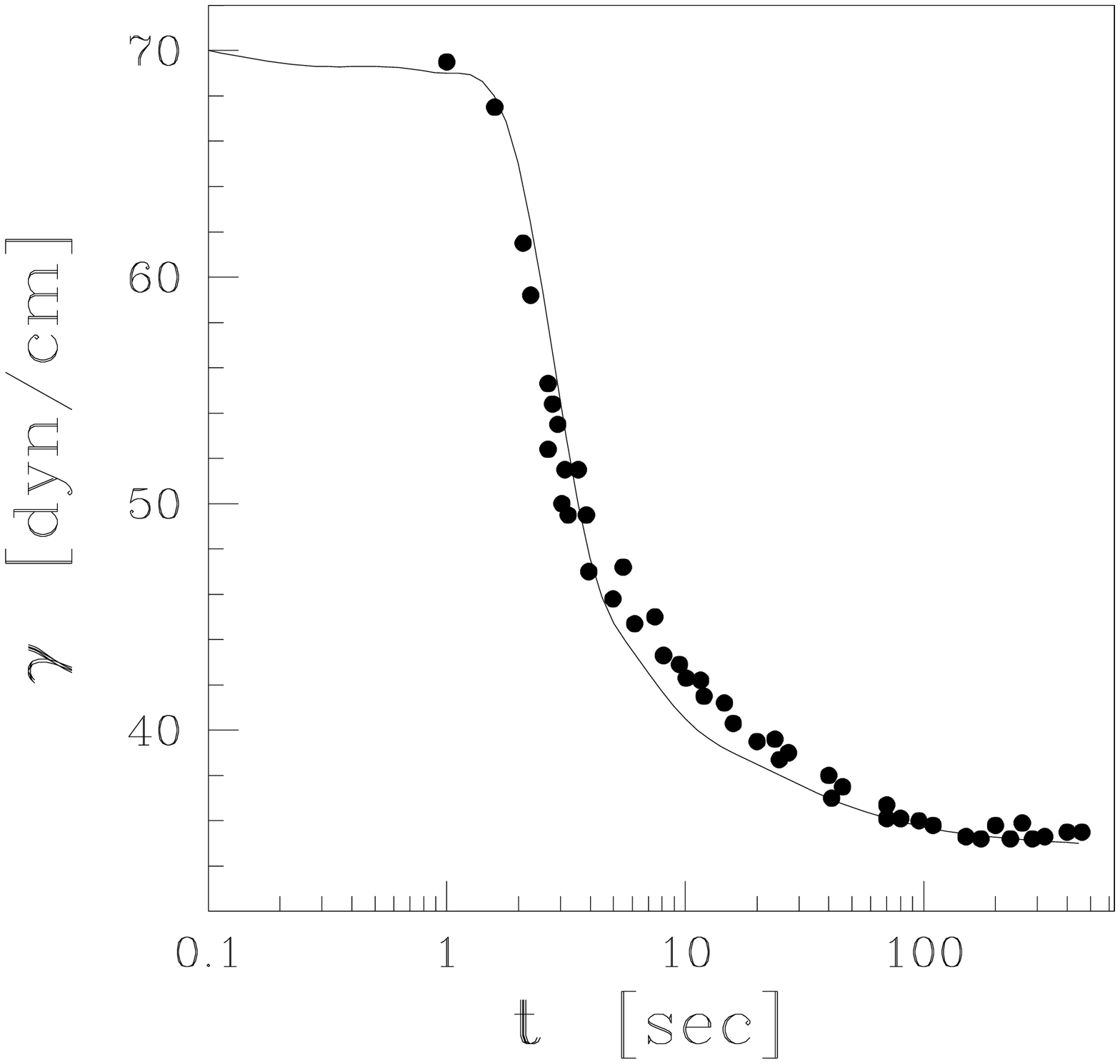}}}
\caption[]{}
\label{fig_numeric_fit}
\end{figure}

Using
the scheme of Eq.~\ref{kinetic_general}
to derive the kinetic
equations,
an ordinary diffusion equation is obtained in the bulk,
\begin{equation}
   \frac{\partial\phi}{\partial t} = D \frac{\partial^2\phi}{\partial x^2},
 \label{diffusion}
\end{equation}
where $D$ is the surfactant diffusion coefficient, assumed constant
in the dilute bulk.
In addition, we get a conservation condition at the sub-surface layer,
\begin{equation}
   \frac{\partial\phi_1}{\partial t} = \frac{D}{a} \left.
        \frac{\partial\phi}{\partial x} \right|_{x=a}
        - \frac{\partial\phi_0}{\partial t},
 \label{dp1dt}
\end{equation}
where $\phi_1\equiv\phi(x\rightarrow 0)$ is the local volume fraction
in the sub-surface layer, to be distinguished from the interfacial
volume fraction, $\phi_0$.
Finally, at the interface itself, we get
\begin{equation}
  \frac{\partial\phi_0}{\partial t} = \frac{D_0}{a^2} \phi_1
          \left[ \ln \frac{\phi_1(1-\phi_0)}{\phi_0} + \frac{\alpha}{T}
          + \frac{\beta\phi_0}{T} \right],
 \label{dp0dt}
\end{equation}
where $D_0$ may differ from $D$.
Applying the Laplace transform with respect to time to
Eqs.~\ref{diffusion} and \ref{dp1dt}, we obtain a relation
similar to that of Ward and Tordai \cite{WT},
\begin{equation}
  \phi_0(t) = \left(\frac{D}{\pi a^2}\right)^{1/2} \left[
          2\phi_{\rm b}\sqrt{t} - \int_0^t \frac {\phi_1(\tau)}
          {\sqrt{t-\tau}} {\rm d}\tau \right]
          + 2\phi_{\rm b} - \phi_1.
 \label{WT}
\end{equation}
The system of two equations, \ref{dp0dt} and \ref{WT},
with appropriate initial
conditions, completely determines the adsorption kinetics
and equilibrium state.
Full solution of the equations can be obtained only numerically.
Several numerical schemes have been proposed for solving the
Ward-Tordai equation
with various boundary conditions \cite{review1,review2,Lin1,ourmixture}.
An example for a numerical solution fitted to experiment is given
in Fig.~\ref{fig_numeric_fit}.
%

\begin{figure}[p]
\centerline{ \resizebox{0.57\textwidth}{!}
{\includegraphics{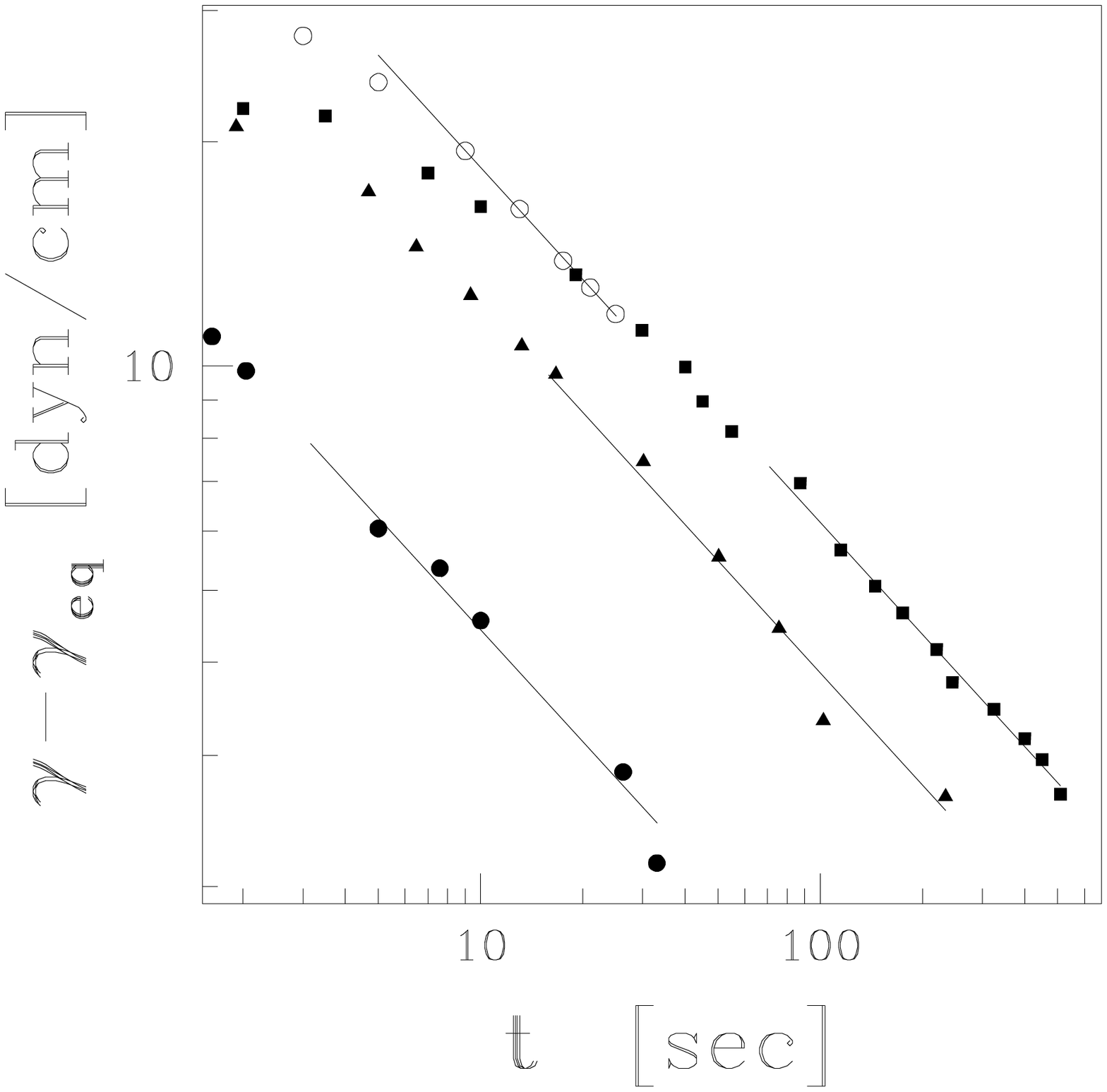}}}
\caption[]{}
\label{fig_DLA}
\end{figure}

Our formalism has led to a diffusive transport
in the bulk (Eq.~\ref{WT}) coupled to
an adsorption mechanism at the interface (Eq.~\ref{dp0dt}).
Let us examine the characteristic time scales
associated with these kinetic equations.
The diffusive transport from the bulk solution (Eq.~\ref{WT})
relaxes like \cite{Hansen}
\begin{equation}
  \phi_1(t\rightarrow\infty)/\phi_{\rm b} \simeq
  1 - (\tau_1/t)^{1/2}, \ \ \ \
  \tau_1 \equiv \frac{a^2}{\pi D} \left(\frac{\phi_{0,{\eq}}}{\phi_{\rm b}}\right)^2,
 \label{asymdiff}
\end{equation}

\noindent
where $\phi_{0,{\eq}}$ denotes the equilibrium surface coverage.
The molecular diffusion time scale, $a^2/D$, is of order
$10^{-9}$~sec, but the factor $\phi_{0,{\eq}}/\phi_{\rm b}$ in
surfactant monolayers
can be very large (typically 10$^5$--10$^6$), so the diffusive
transport to
the interface may require minutes.
On the other hand, the kinetic process at the interface
(Eq.~\ref{dp0dt})
relaxes like
\begin{equation}
  \phi_0(t\rightarrow\infty)/\phi_{0,{\eq}} \simeq
  1 - \rme^{-t/\tau_{\rm k}},
\ \ \ \
  \tau_{\rm k} \equiv \frac{a^2}{D_0} \left(\frac{\phi_{0,{\eq}}}{\phi_{\rm b}}\right)^2
     \rme^{-(\alpha+\beta\phi_{0,{\eq}})/T}.
 \label{asymkin}
\end{equation}
Typical values of $\alpha$ for common surfactants are much larger
than $T$. In the absence of barriers hindering the kinetics at the
interface, $D_0$ is not expected to be drastically smaller than
$D$, and $\tau_{\rm k}$, therefore, is much smaller than $\tau_1$.
In other words, the adsorption of common non-ionic surfactants is
expected to be {\em diffusion-limited}. The asymptotic time
dependence found in Eq.~\ref{asymdiff}
gives a distinct `footprint' for
diffusion-limited adsorption, as demonstrated in
Fig.~\ref{fig_DLA}.

One consequence of a diffusion-limited process
is that the relation between $\phi_0$ and $\phi_1$ is given at
all times by the equilibrium adsorption isotherm
(Eq.~\ref{Frumkin} in our model).
The solution of the adsorption problem in that
case amounts, therefore, to the
solution of the Ward-Tordai equation \ref{WT} with
the adsorption isotherm as a boundary condition.
An exact analytical solution exists only for the simplest, linear
isotherm, $\phi_0\propto\phi_1$ \cite{Sutherland}.
Such an approximation, however, is valid only for low surface coverage
and, hence, not very useful for the description of the entire
adsorption process \cite{ourmixture}.
For more realistic isotherms such as Eq.~\ref{Frumkin}, one has
to resort to numerical techniques, as mentioned above and demonstrated
in Fig.~\ref{fig_numeric_fit}.
Another consequence of a diffusion-limited process, as explained
in Sec.~\ref{sec_framework},
is that the dynamic surface tension,
$\Delta\gamma(t)$, approximately obeys the equilibrium equation of
state \ref{eqstate}.
These results show that the validity of schemes
employed by previous theories is essentially restricted to
diffusion-limited cases.

\begin{figure}[p]
\centerline{
  \resizebox{0.57\textwidth}{!}
  {\includegraphics{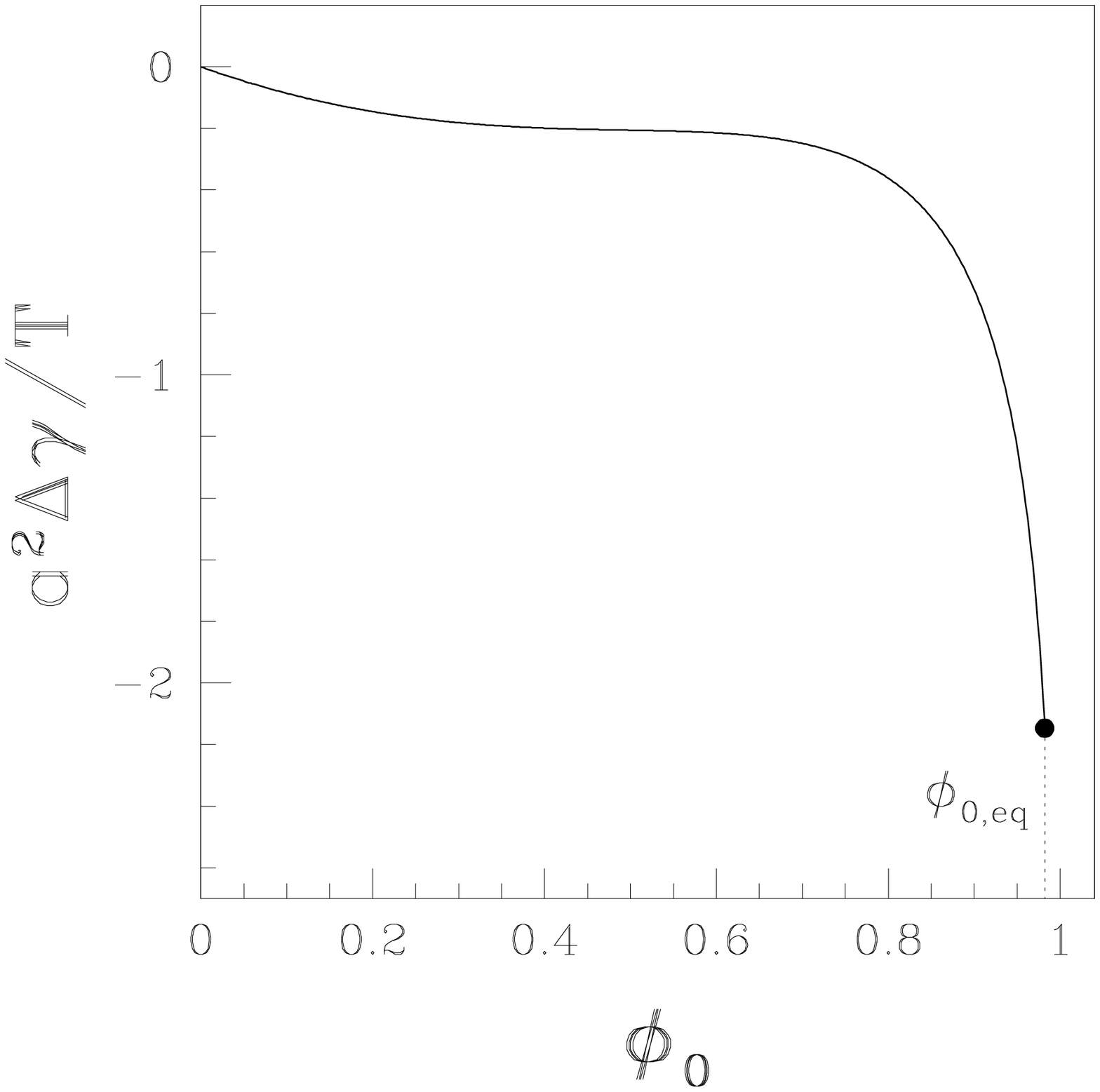}}}
\caption[]{}
\label{fig_st_DLA}
\end{figure}

The dependence defined by the equilibrium equation of state
\ref{eqstate} is depicted in Fig.~\ref{fig_st_DLA}.
As a result of the competition between the entropy and interaction
terms in the equation, the surface tension changes very little
for small surface coverages.
As the coverage increases beyond about $1-(\beta/T)^{-1/2}$, the
surface tension starts decreasing until reaching equilibrium.
This qualitatively explains the shape of dynamic surface tension
curves found in experiments for non-ionic surfactants (see
Fig.~\ref{fig_numeric_fit}).
When the adsorption is not diffusion-limited, this
theoretical approach is no longer applicable, as will be demonstrated
in the ionic case.

In a diffusion-limited process the various physical quantities all
have the asymptotic characteristic $t^{-1/2}$ dependence, similar
to Eq.~\ref{asymdiff}. Yet, the characteristic relaxation times 
$\tau_0$, $\tau_1$ and $\tau_\gamma$, characterizing the 
temporal decay of $\phi_0$, $\phi_1$ and $\gamma$, respectively, 
may differ:
\begin{equation}
   \phi_0/\phi_{0,\eq}\simeq 1-(\tau_0/t)^{1/2},\ \
\phi_1/\phi_{\rm b}\simeq 1-(\tau_1/t)^{1/2},\ \
   \Delta\gamma/\Delta\gamma_{\eq}\simeq 1-(\tau_\gamma/t)^{1/2}.
\label{various_tau}
\end{equation}
Experiments are usually concerned with surface coverage and surface
tension, rather than sub-surface concentration.
Substituting $\phi_1$ of Eq.~\ref{asymdiff} in Eqs.~\ref{Frumkin} 
and \ref{eqstate}, we find
\begin{equation}
  \tau_0 = \left\{ {(1-\phi_{0,\eq})}/{[1-(\beta/T)\phi_{0,\eq}
      (1-\phi_{0,\eq})]} \right\}^2 \tau_1, \ \ \ \
  \tau_\gamma = (\phi_{0,\eq})^2 \tau_1.
\label{tau_gamma}
\end{equation}
Since $\phi_{0,\eq}$ is usually very close to unity, the value of
$\tau_\gamma$ extracted from dynamic surface tension measure\-ments
is practically identical to $\tau_1$ of Eq.~\ref{asymdiff}.
(The possible divergence of $\tau_0$ for $\beta>4T$ is a consequence
of the non-convexity of $f_0$, Eq.~\ref{f0}, for these values of
$\beta$, indicating a transition to a two-phase coexistence.)

\subsection{Short Time Behavior}

In order
to provide a comprehensive
description of the adsorption process, the time dependence
during early stages is of interest as well.
It should be first noted that diffusion-limited behavior cannot
strictly start at $t=0$,
since at that instance the interface and sub-surface layers are
not at equilibrium with
each other.
Assuming a diffusion-limited time dependence of the form
$\phi_0(t) \simeq \mbox{const.} + (t /\tau_1)^{1/2}$ \cite{Hansen},
the $\mbox{const.}$ is found to be roughly equal
to $2\phi_{\rm b}$.
(This can be obtained also from the analytic solution
of the diffusion-limited problem in the linear adsorption limit;
see Ref.~\cite{ourmixture}.)
In other words, only once
the surface coverage has exceeded a value of $2\phi_{\rm b}$,
can one assume a process limited by diffusion.
Prior to the onset of diffusion, a short stage takes place,
when most of the molecules in the
sub-surface layer rapidly adsorb onto the interface.
Only when the sub-surface
layer becomes almost completely depleted, do molecules from the bulk
start migrating towards the interface by a diffusive
mechanism.
To address these very early time stages, the interfacial kinetics
must be considered explicitly.
Assuming that the bulk solution is still at its initial
equilibrium state,
unperturbed by the presence of the interface,
the leading time behavior of the surface coverage is found
from Eq.~\ref{dp0dt} to be linear,
\begin{equation}
\phi_0 (t \rightarrow 0) \simeq \phi_{\rm b}[1 + (D_0/a^2)
  (\alpha/T) t].
\end{equation}
A surface coverage of $2\phi_{\rm b}$ is thus attained after
a period of about $a^2 T/(D_0\alpha)$.
This time scale is typically extremely short (smaller than
microseconds),
unless the adsorption is hindered by barriers making $D_0$
drastically smaller than $D$.
Hence, these very early time stages
are usually of no experimental interest, and the measured initial
time dependence is of a diffusion-limited form, \ie proportional
to $t^{1/2}$.

\section{Non-Ionic Mixtures}
\setcounter{equation}{0}

In the next example we study the adsorption from a mixture of two
non-ionic surfactants \cite{ourmixture}. Surfactant mixtures are
used in numerous industrial applications, and are also encountered
in many systems because of the presence of surface-active
impurities. The {\em equilibrium} behavior of mixed surfactant
solutions was studied in detail in previous works
\cite{Lucassen}--\cite{Blank}. One of the important results, both
theoretically and from the application point of view, is the
ability to relate the mixed-surfactant behavior with that of the
better understood, single-surfactant one. One of our aims is to
predict the mixture {\em kinetics} from the behavior of the single
surfactants. A particularly interesting question is whether mixing
several species would lead in certain cases to a significant
difference in the kinetics as compared to the single-surfactant
systems.

We consider two surfactants denoted A and B. The same notation as
in the previous section is used, except for the following
modifications. We use $\phi$ to denote volume fraction of
surfactant A and $\psi$ for that of surfactant B. Parameters
characterizing the two surfactants, such as $\alpha$, $\beta$, $D$
etc., are distinguished by subscripts A and B. The subscripts
$0,1,{\rm b}$ are used, as in the previous section, to denote
different positions in the solution (interface, sub-interface
and bulk, respectively).

The excess free energy of Eq.~\ref{Dg_general}
is written in the mixture case as
\begin{equation}
   \Delta\gamma[\phi, \psi ] = \int_{0}^{\infty} \left\{
 \Delta f [\phi(x) ] + \Delta f [ \psi
(x) ]\right\} \rmd x ~+~  f_{0} (\phi_{0} , \psi_{0} ).
\label{a13}
\end{equation}
Since the solution is dilute, the two species are assumed to be
uncorrelated in the bulk. The bulk free energy is taken,
therefore, as a sum of single-surfactant contributions, given by
Eq.~\ref{Df}. The surfactant molecular size, $a$, is assumed
to have the same value for both species, on account of simplicity
\cite{dif_a}. At the interface, due to the high surface coverage,
coupling terms must be considered,
\begin{eqnarray}
 f_{0} (\phi_{0} , \psi_{0} ) &=&
  \{ T[ \phi_{0}
 \ln \phi_{0} + \psi_{0} \ln \psi_{0}
   + \eta_{0} \ln \eta_{0}]
 - (\alpha_{\rm A}+ \mu_{1,{\rm A}})\phi_{0} -
(\alpha_{\rm B} + \mu_{1,{\rm B}})\psi_{0}
\nonumber \\
 & & ~ -(\beta_{\rm A}/2)  {\phi_{0} }^{2} -
(\beta_{\rm B}/ 2) { \psi_{0} }^{2} -
\chi \phi_{0} \psi_{0} \} / a^2,
\label{a14}
\end{eqnarray}
where additional interaction between different surfactants has been
introduced, having a characteristic energy $\chi$.
Note that this is a {\rm tertiary} system
(two solutes in a solvent), requiring three parameters for a
complete description of the interactions (in our case
$\beta_{\rm A}$, $\beta_{\rm B}$ and $\chi$).
For brevity we use $\eta_{0} \equiv 1 - \phi_{0} - \psi_{0}$ as
the surface coverage of the solvent (water).

The uncorrelated contributions of the two species, $\Delta f(\phi)$
and
$\Delta f(\psi)$, result in decoupled equilibrium and kinetic
equations in the bulk.
Any correlation between the surfactants in this model originates,
therefore, from interfacial interactions.

\subsection{Equilibrium Relations}

Following
the scheme of Eq.~\ref{equilibrium_general} to derive
equilibrium relations,
two uniform profiles are obtained in the bulk,
$ \phi (x>0) \equiv \phi_{\rm b}$  and
$ \psi (x>0) \equiv \psi_{\rm b}$, and
at the interface we get a Frumkin adsorption
isotherm, generalized for the A/B mixture case:
\begin{equation}
\phi_{0} = \frac{ \phi_{\rm b} ( 1 - \psi_{0} ) }{ \phi_{\rm b}
   + \rme^{  -( \alpha_{\rm A} + \beta_{\rm A} \phi_{0} + \chi
     \psi_{0} ) /T } },
\ \ \ \
\psi_{0} = \frac{ \psi_{\rm b} ( 1 - \phi_{0} ) }{ \psi_{\rm b}
   + \rme^{  -( \alpha_{\rm B} + \beta_{\rm B} \psi_{0} + \chi
     \phi_{0} )/T  } }
\label{a18}
\end{equation}
The adsorption of species A depends on species B
through the entropy of mixing (steric effect) and
surfactant--surfactant interactions.
Finally, the equilibrium equation of state,
$ \Delta \gamma = \Delta \gamma ( \phi_0,\psi_0) $,
takes the form
\begin{equation}
   \Delta\gamma = [ T  \ln \eta_{0} +
(\beta_{\rm A}/ 2)  {\phi_{0}}^{2} +
(\beta_{\rm B}/2) {\psi_{0}}^{2} + \chi
\phi_{0} \psi_{0}  ]/a^2.
\label{a19}
\end{equation}

\subsection{Kinetic Equations}

Applying
the scheme of Eq.~\ref{kinetic_general} to the current
free-energy functional yields two single-surfactant diffusion
equations like Eq.~\ref{diffusion} for the two species.
Consequently, two decoupled Ward-Tordai equations like
Eq.~\ref{WT} are obtained as well.
At the interface, however, the two species are correlated
and the scheme yields two coupled kinetic equations:
\begin{eqnarray}
  \frac{ \partial \phi_{0} } {\partial t}  &=&
\frac{D_{\rm A}}{ a^2} \phi_{1}  \left[ \ln \left(
\frac {\phi_{1}\eta_{0} }{ \phi_{0} }   \right) +
\frac{\alpha_{\rm A}} {T} +
\frac{\beta_{\rm A}\phi_{0}} {T} +
\frac{\chi\psi_{0}} {T}  \right] \nonumber \\
%
   \frac{\partial \psi_{0} } {\partial t} &=&
 \frac{D_{\rm B}}{a^{2}} \psi_{1} \left[ \ln \left(
\frac {\psi_{1}\eta_{0} } { \psi_{0} }  \right) + 
\frac {\alpha_{\rm B}} {T} +
\frac {\beta_{\rm B} \psi_{0}} {T} + 
\frac {\chi \phi_{0}} {T}
\right].
\label{a12}
\end{eqnarray}
As can be seen from Eqs. \ref{a12},
the coupling between the kinetics of the two species
arises from an interaction term
as well as from an entropic one ({\it via} $\eta_0$).
The system of four equations (two Ward-Tordai equations like
Eq.~\ref{WT} and the two equations \ref{a12}), with the
appropriate initial conditions, completely determines the
mixture kinetics and equilibrium state.

\begin{figure}[p]
\centerline{
  \resizebox{.57\textwidth}{!}
  {\includegraphics{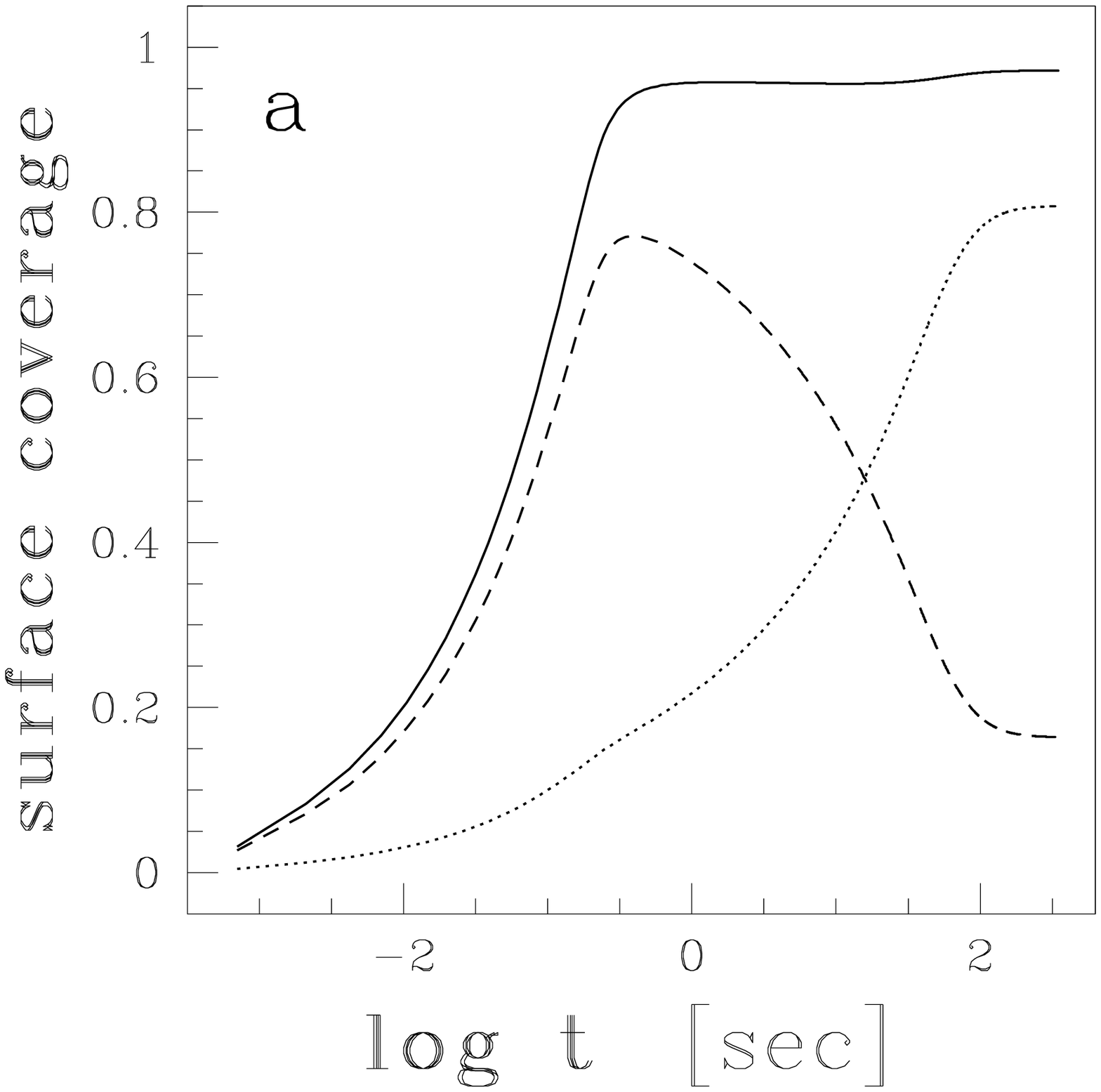}}
  \resizebox{.57\textwidth}{!}
  {\includegraphics{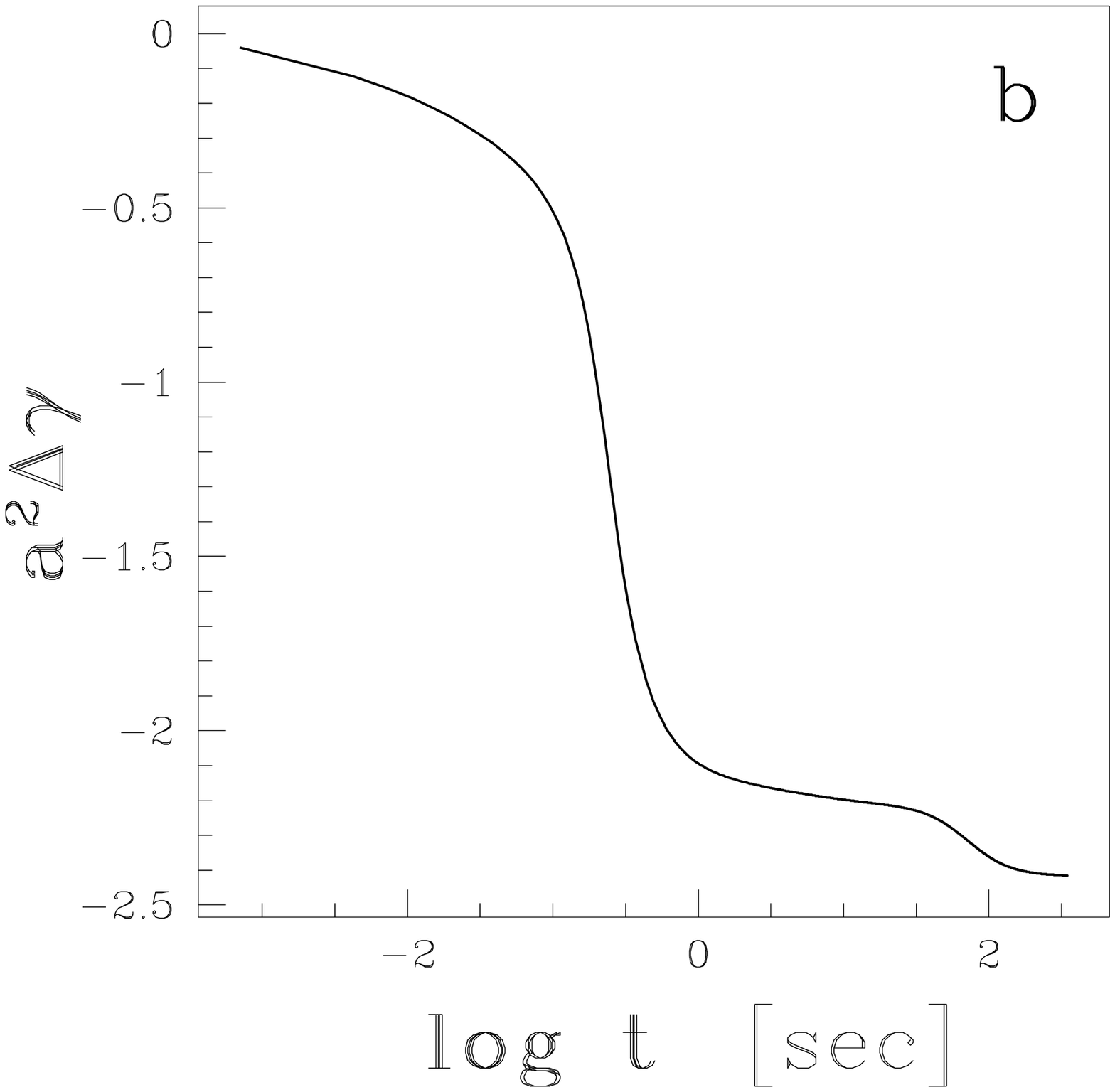}}}
\caption[]{}
\label{fig_numeric_mix}
\end{figure}

The set of equations can be fully solved numerically.
We generalized the recursive scheme of
Miller {\em et al.} \cite{review2}
to a surfactant mixture
having time-dependent boundary conditions.
An example for the resulting time dependence of the various
quantities is given in Fig.~\ref{fig_numeric_mix}.
The mixture parameters were specifically chosen
to show the interesting case of competition
between the two species. While surfactant B diffuses more
rapidly and is more abundant at the interface during the initial
stages of adsorption, surfactant A has a higher
surface affinity and dominates the later stages.
We note that due to this competition,
not only does surfactant A take over the adsorption at the
later time stages, but it also forces surfactant B to desorb from
the interface.
As shown in Fig.~\ref{fig_numeric_mix}b, the competition between
surfactants leads to a more complex decrease of
the surface tension at intermediate times.

As in the previous section, we are interested in the
characteristic time scales of the mixture kinetics.
Assuming a diffusion-limited adsorption, the relaxation time scales
of the two sub-surface concentrations, $\tau_{1,{\rm A}}$ and
$\tau_{1,{\rm B}}$, are found to be identical to the
single-surfactant result, Eq.~\ref{asymdiff}.
They are still inter-dependent, however, since the presence
of each species changes the equilibrium surface coverage of the
other.
The coupling appears more explicitly in the time scales of the
surface coverage, $\tau_0$, and surface tension, $\tau_\gamma$.
Two coupled linear equations are obtained for $\tau_{0,{\rm A}}$
and $\tau_{0,{\rm B}}$,
\begin{eqnarray}
\eta_0 \sqrt{ \tau_{1,{\rm A}} } &=& [ 1 - \psi_0 - (\beta_{\rm
A}/T) \phi_0 \eta_0  ] \sqrt{ \tau_{0,{\rm A}} }
 ~~+~~
\psi_0 [ 1 -  (\chi/T) \eta_0] \sqrt{ \tau_{0,{\rm B}} }
\nonumber\\
\eta_0 \sqrt{ \tau_{1,{\rm B}} } &=& [ 1 - \phi_0 -
(\beta_{\rm B}/T) \psi_0 \eta_0 ] \sqrt{ \tau_{0,{\rm B}} } ~~+~~
 \phi_0 [ 1 - (\chi/T) \eta_0 ] \sqrt{ \tau_{0,{\rm A}} },
\label{tau_1_2}
\end{eqnarray}
where the subscript `eq' has been omitted for brevity.
The expression for $\tau_\gamma$ also combines contributions from both
species,
\begin{eqnarray}
-a^2\Delta\gamma \sqrt{ \tau_{\gamma} } &=& [ \phi_{0} /  \eta_{0}
- (\beta_{\rm A}/T) \phi_{0}^{2} - (\chi/T) \phi_{0} \psi_{0} ]
\sqrt{ \tau_{0,{\rm A}} } \ \ + \nonumber \\
 & & [ \psi_{0} / \eta_{0}  -
 (\beta_{\rm B}/T) \psi_{0}^{2} - (\chi/T) \phi_{0}
\psi_{0} ] \sqrt{ \tau_{0,{\rm B}} }.
\label{tau_gamma2}
\end{eqnarray}

If we `turn off' interactions ($\beta_{\rm A}=\beta_{\rm
B}=\chi=0$), Eq.~\ref{tau_gamma2} is reduced to a simple
expression, relating $\tau_{\gamma}$ of the mixture with those of
each species separately, $\bar{\tau}_{\gamma,{\rm A}}$ and
$\bar{\tau}_{\gamma,{\rm B}}$ (given each by
Eq.~\ref{tau_gamma}),
\begin{equation}
\label{mix}%
\Delta\gamma\sqrt{ \tau_{\gamma} } = \Delta\bar{\gamma}_{\rm A}
\left( \phi_{0} / \bar{\phi_{0}} \right)^{2} \sqrt{
\bar{\tau}_{\gamma,{\rm A}} } + \Delta\bar{\gamma}_{\rm B} \left(
\psi_{0} / \bar{\psi_{0}}  \right)^{2} \sqrt{
\bar{\tau}_{\gamma,{\rm B}} } \ ,
\end{equation}
where $\bar{\phi_{0}}$ and $\bar{\psi_{0}}$ denote the surface
coverages of the single-surfactant systems and
$\Delta\bar{\gamma}_{\rm A}$, $\Delta\bar{\gamma}_{\rm B}$ the
corresponding changes in equilibrium surface tension. Equation
\ref{mix} is a `weighting formula' for relating
the time scale of surface tension
relaxation in the mixture with those of
its individual constituents. It provides, therefore,
a convenient tool for predicting the behavior of
multi-component surfactant mixtures, based on single-surfactant
data. In Table~1 the predicted
$\tau_\gamma$ of Eq.~\ref{mix} is compared
with experimental results
obtained by Fainerman and Miller \cite{Fainerman2}
for a sequence
of Triton~X mixtures. Based on single-surfactant values and
equilibrium isotherms for the mixture, the two terms of
Eq.~\ref{mix} are calculated separately. The agreement between
theory and experiment is quite good, although experiments were
limited to cases having one species dominating the adsorption. The
last entry in the table corresponds to a mixture of Triton X-405
and Triton X-165. Here the predicted $\tau_\gamma$ deviates from
the experimental one by 33\%. Equilibrium measurements on this
mixture reveal an {\em increase} in X-165 coverage upon addition
of X-405 \cite{Fainerman2},
indicating strong interfacial
interactions between the species. The deviation in the predicted
kinetics in Table~1 is attributed to those interactions, which are
not taken into account by Eq.~\ref{mix}.
(It is possible to treat also the general case, including
interactions, by using the full equations
\ref{tau_1_2} and \ref{tau_gamma2} instead of the simplified
equation \ref{mix}. Such a procedure, however, involves
three additional fitting parameters --- $\beta_{\rm A}$,
$\beta_{\rm B}$ and $\chi$.)

\subsection{Kinetically Limited Adsorption}

Although
most non-ionic surfactants undergo a
diffusion-limited process, as was discussed in the previous
section, the adsorption of certain surfactants
is found to be kinetically limited due to adsorption barriers.
It is of interest, therefore, to examine the mixture kinetics
in the kinetically limited case.
The equations governing such a process are the two coupled
interfacial equations \ref{a12}.
Linearizing about the equilibrium state, $\phi_{0,\eq}$ and
$\psi_{0,\eq}$,
two time scales denoted $\tau_{+}$ and $\tau_{-}$ emerge
($ \tau_{-} > \tau_{+}$).
These collective time scales correspond to the kinetics of a
certain combination of surfactant coverages,
\begin{equation}
 C_1 \Delta \phi_0 + C_2 \Delta \psi_0 \sim
 \rme^{ - t/\tau_{-} }, \ \ \ \
 C_3 \Delta \phi_0 +  C_4 \Delta \psi_0 \sim
 \rme^{ - t/\tau_{+} },
\label{longKLAmix}
\end{equation}
where $\Delta \phi_0 \equiv \phi_0-\phi_{0,\eq}$,
$\Delta \psi_0 \equiv \psi_0-\psi_{0,\eq}$,  and
$C_1 \ldots C_4$ are constants.
Since  $ \tau_{-} > \tau_{+} $, it is $ \tau_{-}$ which
limits the kinetics of the system.

In the simple case of no surface interactions
($ \beta_{\rm A} = \beta_{\rm B} = \chi = 0$), the expressions
for $ \tau_{\pm} $ are
\begin{equation}
{ 2 }/{ \tau_{\pm} } =
{ (1 - \psi_{0}) }/{ \tau_{\rm A} } +
{ (1 - \phi_{0}) }/{ \tau_{\rm B} }
\pm  \sqrt{ \left[ { (1 - \psi_{0}) }/{ \tau_{\rm A} } +
{ (1 - \phi_{0}) }/{ \tau_{\rm B} } \right]^2
- {  4\eta_0 }/{ (\tau_{\rm A} \tau_{\rm B}) }  },
\label{TauK}
\end{equation}
where $ \tau_{\rm A} $ and $ \tau_{\rm B} $ are the time scales of
the single-surfactant case, formulated in Eq.~\ref{asymkin},
yet with $\phi_{0}$ and $\psi_{0}$ of the {\it mixture}.
The behavior of the mixed system combines the single-surfactant
kinetics in a complicated manner.
We can gain some insight on this coupling
by considering two simple cases.
In the limit where the interfacial kinetics of surfactant A is
much slower than that of B,
$\tau_{\rm A} \gg \tau_{\rm B}, $ Eqs.~\ref{longKLAmix}
and \ref{TauK} are simplified to
\begin{eqnarray}
( 1 - \phi_{0,\eq}) \Delta \phi_0 -  \psi_{0,\eq}  \Delta \psi_0
   &\sim& \rme^{ -t/\tau_{-} }, \ \  \tau_{-} =\tau_{\rm A}
( 1- \phi_0 )  / \eta_0
   \nonumber \\
\Delta \psi_0  &\sim& \rme^{ -t/\tau_{+} },\ \  \tau_{+} =
\tau_{\rm B}/( 1 - \phi_0 )
\label{Ta_ll_Tb}
\end{eqnarray}
In the other limit, where the two species have similar
time scales,
$\tau_{\rm A} \simeq \tau_{\rm B}$, we get
\begin{eqnarray}
\Delta \phi_0  -  \Delta \psi_0
   &\sim& \rme^{ -t/\tau_{-} },\ \  \tau_{-} =  \tau_{\rm A} /
\eta_0
   \nonumber \\
\phi_{0,\eq}  \Delta \phi_0 + \psi_{0,\eq} \Delta \psi_0
   &\sim& \rme^{ -t/\tau_{+} },\ \  \tau_{+} = \tau_{\rm A}
\label{Ta_eq_Tb}
\end{eqnarray}
The factor $1/\eta_0$ in $\tau_{-}$ is quite interesting.
Since the equilibrium surface coverage of the solvent, $\eta_0$, is
usually very small in surfactant systems, this factor implies that
the coupling in a surfactant mixture undergoing kinetically
limited adsorption may lead to a
significant reduction in adsorption rate. In this regime
the mixture behavior may differ considerably from that of its
individual constituents.
Due to the relatively  large factor of $1/\eta_0$,  the
time scale of interfacial kinetics may exceed the diffusive one and
the adsorption would then become kinetically limited.

\section{Ionic Surfactants}
\setcounter{equation}{0}

We turn to the more complicated, yet important problem of ionic
surfactant adsorption \cite{JPC}, and start with the salt-free
case where strong electrostatic interactions are present. In
Fig.~\ref{fig_plat_exp} we have reproduced experimental results 
reported by
Bonfillon-Colin {\it et al.} \cite{Langevin1,Langevin2} and by Hua
and Rosen \cite{HuaDESS}. The dynamic surface tension of the
investigated ionic salt-free solutions exhibits much longer
kinetics and richer behavior than in common non-ionic systems. A
few theoretical models were suggested for the problem of ionic
surfactant adsorption \cite{Dukhin}--\cite{Radke}, yet none of
them could produce such dynamic surface tension curves. Moreover,
it is rather evident that a theoretical scheme for non-ionic surfactants, 
such as the one discussed in the previous sections,
cannot fit the ionic results. On the other hand, as can be seen in
Fig.~\ref{fig_plat_exp}, 
addition of salt to the solution leads to a very similar behavior,
as compared with the non-ionic case. It is thus inferred that the
different kinetics observed for the salt-free solutions results
from strong electrostatic interactions, which are screened upon
addition of salt. Let us now study this effect in more detail. We
follow the same line presented in the previous sections while
adding appropriate terms to account for the additional
interactions.

The free energy in the current
case is written as a functional of three degrees of freedom:
the surfactant profile, $\phi^+(x,t)$
(we arbitrarily take the surfactant
as the positive ion), the counterion
profile, $\phi^-(x,t)$, and a mean electric potential,
$\Psi(x,t)$,
\begin{eqnarray}
  \Delta \gamma [\phi^+,\phi^-,\Psi] &=& \int_0^\infty
      [ \Delta f(\phi^+) + \Delta f(\phi^-) +
   f_{\rm el}(\phi^+,\phi^-,\Psi) ]
   {\rm d}x \nonumber \\
      & & +~ f_0(\phi^+_0) ~+~ f_{\rm el,0}(\phi^+_0,\Psi_0).
 \label{Dg2}
\end{eqnarray}
The bulk contributions coming from the two profiles,
$\Delta f^\pm$, contain the same terms as in Eq.~\ref{Df}
of the non-ionic case.
The interfacial contribution, $f_0$, is identical to
Eq.~\ref{f0}
and is taken as a function of the surfactant coverage alone,
assuming that the counterions are surface-inactive.
In addition, electrostatic contributions are introduced in the
bulk free energy as well as in the interfacial one, accounting
for interactions between the ions and the electric field and the
energy associated with the field itself,
\begin{eqnarray}
        f_{\rm el} &=& e \left[ {\phi^+}/{(a^+)^3} -
  {\phi^-}/{(a^-)^3}
           \right] \Psi - ({\varepsilon}/{8\pi})
  \left( {\partial\Psi}/{\partial x} \right)^2
        \label{fel} \\
        f_{\rm el,0} &=& [{e}/{(a^+)^2}] \phi^+_0\Psi_0,
\label{fel0}
\end{eqnarray}
where $a^\pm$ are the molecular sizes of the two ions, $e$ the
electronic charge and $\varepsilon\simeq 80$ the dielectric
constant of water.
For simplicity we have restricted ourselves to fully ionized,
monovalent ions, implying that
$\phi^+_{\rm b}/(a^+)^3=\phi^-_{\rm b}/(a^-)^3= c_{\rm b}$,
$c_{\rm b}$ being the bulk concentration.
Ions in solution, apart from interacting with each other, are
subject to
an additional repulsion from the interface due to `image-charge'
effects \cite{Onsager}.
It can be shown, however, that those effects become negligible
in our case as soon as the surface coverage exceeds about 2 percents
\cite{JPC}.

\subsection{Equilibrium Relations}

Employing
the same scheme of Eq.~\ref{equilibrium_general},
the variation with respect to $\phi^\pm(x)$ yields the Boltzmann
ion profiles,
\begin{equation}
  \phi^\pm(x>0) = \phi^\pm_{\rm b} \rme^{\mp e\Psi(x)/T},
\label{Boltzmann}
\end{equation}
with respect to $\Psi(x)$ --- the Poisson equation,
\begin{equation}
  {\partial^2\Psi}/{\partial x^2} = -({4\pi e}/{\varepsilon})
  \left[ {\phi^+}/{(a^+)^3} - {\phi^-}/{(a^-)^3} \right],
 \label{Poisson}
\end{equation}
with respect to $\Psi_0$ --- the electrostatic boundary
condition,
\begin{equation}
  \left. {\partial\Psi}/{\partial x} \right|_{x=0} =
       -[{4\pi e}/{\varepsilon (a^+)^2}] \phi^+_0,
 \label{neutral}
\end{equation}
and, finally, the variation with respect to $\phi^+_0$ recovers the
Davies adsorption isotherm \cite{Davies},
\begin{equation}
  \phi^+_0 =  \frac{\phi^+_{\rm b}} {\phi^+_{\rm b} +
           \rme^{-(\alpha+\beta\phi^+_0-e\Psi_0)/T}}.
 \label{Davies}
\end{equation}
Combining Eqs.~\ref{Boltzmann} and \ref{Poisson} leads to
the well-known Poisson-Boltzmann equation
for the equilibrium double-layer potential \cite{VO,AndelmanES},
\begin{equation}
  {\partial^2\Psi}/{\partial x^2} = ({8\pi e c_{\rm b}}/{\varepsilon})
           \sinh ({e\Psi}/{T}),
 \label{PB}
\end{equation}
By means of the Poisson-Boltzmann equation, the Davies isotherm
\ref{Davies} can be re-expressed as
\begin{equation}
  \phi^+_0 =  \frac{\phi^+_{\rm b}} 
  {\phi^+_{\rm b} + [ b\phi^+_0 + \sqrt{
           (b\phi^+_0)^2+1} ]^2
           \rme^{-(\alpha+\beta\phi^+_0)/T}},
 \label{Davies2}
\end{equation}
where $b \equiv  a^+/(4\phi^+_{\rm b}\lambda)$, and $\lambda
\equiv (8\pi c_{\rm b} e^2/\varepsilon T)^{-1/2}$ is the
Debye-H\"{u}ckel screening length \cite{DH}. The equilibrium
equation of state, relating surface tension and surface coverage,
is
\begin{equation}
  \Delta\gamma = \{ T\ln(1-\phi^+_0) +
               (\beta/2) (\phi^+_0)^2 - (2T/b)
               [\sqrt{(b\phi^+_0)^2+1} - 1] \} / (a^+)^2.
 \label{eqstate2}
\end{equation}
For weak fields the electrostatic correction to the equation of
state (cf.\ Eq.~\ref{eqstate})
is quadratic in the coverage, thus merely modifying the
lateral interaction term, whereas for strong fields it becomes
linear in the coverage.

\subsection{Kinetic Equations}

Applying
the same scheme of Eq.~\ref{kinetic_general} to the
current case yields in the bulk the
Smoluchowski diffusion equations,
\begin{equation}
  \frac{\partial\phi^\pm}{\partial t} = D^\pm \frac{\partial}{\partial x}
      \left( \frac{\partial\phi^\pm}{\partial x} \pm \frac{e}{T}
      \phi^\pm \frac{\partial\Psi}{\partial x} \right),
\label{diffusion2}
\end{equation}
where $D^\pm$ are the diffusion coefficients of the two ions, assumed
to be constant in the dilute bulk.
At the sub-surface we find
\begin{equation}
  \frac{\partial\phi^\pm_1}{\partial t} = \frac{D^\pm}{a^\pm} \left(
        \left.\frac{\partial\phi^\pm}{\partial x}\right|_{x=a^\pm} \pm
        \frac{e}{T} \phi^\pm_1 \left.\frac{\partial\Psi}{\partial x}
        \right|_{x=a^\pm} \right) - \frac{\partial\phi^\pm_0}{\partial t},
\label{dp1dt2}
\end{equation}
and, finally, at the interface itself,
\begin{equation}
  \frac{\partial\phi^+_0}{\partial t} =
          \frac{D^+_0}{(a^+)^2} \phi^+_1
          \left[ \ln \frac{\phi^+_1(1-\phi^+_0)}{\phi^+_0} +
          \frac{\alpha}{T} + \left( \frac{\beta}{T}
          - \frac{4\pi l}{a^+} \right) \phi^+_0 \right],
 \label{dp0dt2}
\end{equation}
where the diffusion coefficient at the interface, $D^+_0$,
may differ from its value in the bulk.
The electrostatic boundary condition, Eq.~\ref{neutral},
has been used in Eq.~\ref{dp0dt2}
to replace an electrostatic barrier term, $e(\Psi_0-\Psi_1)/T$,
with the approximate term $(4\pi l / a^+) \phi^+_0$, where
$l \equiv e^2/\varepsilon T$
is the Bjerrum length (about 7 \AA \ for water at room
temperature).

Neglecting electrodynamic effects, the Poisson equation holds
out of equilibrium as well.
The kinetic equations just derived, along with the Poisson equation
\ref{Poisson},
the boundary condition of Eq.~\ref{neutral},
another boundary condition for
the counterion profile (\eg $\phi^-_0(t)=0$), and
appropriate initial conditions, together determine the kinetics and
equilibrium state of the adsorption problem.
This set of equations can be fully solved only numerically
(a similar set was solved in Ref.~\cite{Radke}).

The relaxation in the bulk solution, accounted for by the
Smoluchowski equations \ref{diffusion2}, has the characteristic
time scale
$
  \tau_{\rm e} = \lambda^2/D
$,
where $D$ is an effective ambipolar diffusion coefficient \cite{LL}.
This time scale is typically very short (of the order of
microseconds),
\ie electrostatic interactions make the bulk relaxation much
faster than in the non-ionic case.
The relaxation at the interface (Eq.~\ref{dp0dt2})
has an asymptotic exponential form like Eq.~\ref{asymkin}.
It is dramatically slowed down, however, by electrostatic
repulsion, having a time scale of
\[
  \tau_{\rm k} = \tau_{\rm k}^{(0)} \exp [e(\Psi_{0}+\Psi_{1})/T]
           \simeq \tau_{\rm k}^{(0)} [ (a^+/2\lambda)
  (\phi^+_{0,\eq}/\phi^+_{\rm b})]^4
                \exp [ -(4\pi l/a^+) \phi^+_{0,\eq}],
\]
where $\tau_{\rm k}^{(0)}$ denotes the kinetic time scale in the
absence of electrostatics (Eq.~\ref{asymkin}).
In salt-free surfactant solutions the surface potential reaches
values significantly larger than $T/e$, and, hence, the interfacial
relaxation is by orders of magnitude slower than in the non-ionic
case.

The conclusion is that ionic surfactants in
salt-free solutions should, in many cases, undergo
{\em kinetically limited adsorption}.
Due to the strong electrostatic repulsion, unlike the non-ionic case,
the adsorption can become kinetically
limited even if the diffusion coefficient at the interface is not
significantly larger than that in the bulk.
Indeed, dynamic surface tension curves of such solutions exhibit
an exponential asymptotic time dependence, rather than the
diffusive $t^{-1/2}$ behavior, as is demonstrated in
Fig.~\ref{fig_exponent}.

The scheme employed for non-ionic surfactants, focusing on the
diffusive transport inside the solution, is no longer valid.
By contrast, the diffusive relaxation in the bulk is practically
immediate and we should concentrate on the interfacial kinetics,
Eq.~\ref{dp0dt2}.
In this case the sub-surface volume fraction, $\phi^+_1$, obeys
the Boltzmann law (Eq.~\ref{Boltzmann}) 
rather than the Davies adsorption isotherm (Eq.~\ref{Davies}),
and the electric potential is given by the Poisson-Boltzmann theory.
Using these results, Eq.~\ref{dp0dt2} can be expressed as a
function of the surface coverage alone,
\begin{equation}
  \frac{\partial\phi^+_0}{\partial t} = 
\frac{D^+_0\phi^+_{\rm b}}{(a^+)^2} 
  \frac{\exp [(4\pi l/a^+)\phi^+_0]} {[b\phi^+_0+\sqrt{(b\phi^+_0)^2+1}]^2}
       \left\{ \ln \left[ \frac{\phi^+_{\rm b}(1-\phi^+_0)}{\phi^+_0} \right] +
      \frac{\alpha}{T} + \frac{\beta\phi^+_0}{T} - 2\sinh^{-1}(b\phi^+_0) \right\},
 \label{dp0dt2_kin}
\end{equation}
thus reducing the problem to a single integration.

Not only does the scheme for solving the kinetic equations
differ from the non-ionic case, but also the way to calculate the
dynamic surface tension has to change.
In kinetically limited adsorption the variation of the free energy
with respect to the surface coverage does not vanish, and,
consequently,
the equation of state \ref{eqstate2} is strictly invalid out of
equilibrium.
The expression for the dynamic surface tension in the kinetically
limited case can be derived from the general functional of
Eq.~\ref{Dg2}
by assuming quasi-equilibrium inside the bulk solution
(\ie using Boltzmann profiles and the Poisson-Boltzmann equation):
\begin{eqnarray}
  \Delta\gamma[\phi^+_0(t)] &=& \{ T [ \phi^+_0\ln(\phi^+_0/\phi^+_{\rm b})
        + (1-\phi^+_0)\ln(1-\phi^+_0) ] - \alpha\phi^+_0 -
  (\beta/2)(\phi_0^+)^2
  \nonumber \\
      & & +~ 2T [ \phi^+_0\sinh^{-1}(b\phi^+_0) -
  (\sqrt{(b\phi^+_0)^2+1} - 1)/b ]
                \} / (a^+)^2.
\label{Dgkinetic}
\end{eqnarray}

Assuming high surface potentials ($b\phi^+_0\gg 1$), the function
defined in
Eq.~\ref{Dgkinetic} becomes non-convex for
$\beta/T > 2(2+\sqrt{3}) \simeq 7.5$, as demonstrated in
Fig.~\ref{fig_plat_theo}a.
In such cases an unusual time dependence for the dynamic surface
tension results (Fig.~\ref{fig_plat_theo}b).
We thus infer that the shape of experimental dynamic surface tension
curves, such as those presented in Fig.~\ref{fig_plat_exp},
is a consequence of a
kinetically limited adsorption brought about
by strong electrostatic interactions.
Physically, the non-convexity
implies a sort of two-phase coexistence, suggesting the
following scenario.
As the surface coverage increases, the system reaches a local
free-energy minimum leading to a pause in the adsorption
(the intermediate plateau of the experimental curves).
This metastable state lasts until domains of the denser,
global-minimum phase are nucleated, resulting in further increase
in coverage and a corresponding decrease in surface tension.
In Fig.~\ref{fig_plat_theo} we have exploited a special set of
parameters in order to demonstrate the effect of non-convexity
within our current formalism.
A complete treatment of the scenario described above, however,
cannot be presented within such a formalism, since it inevitably
leads to a monotonically decreasing free energy as a function of
time, and hence, cannot account for nucleation  \cite{Langer}.

A value of $\beta > 7.5T$ required for non-convexity is somewhat
large compared to the typical lateral attraction between
surfactant molecules. Throughout the above calculations we have
assumed that no counterions are adsorbed at the interfacial layer.
It can be shown that the presence of a small amount of counterions
at the interface introduces a correction to the free energy which
is quadratic in the surfactant coverage, \ie leading to an
effective increase in lateral attraction \cite{JPC}. The increase
in $\beta$ due to the counterions turns out to be $[2\pi l
a^-/(a^+)^2]T$, which may amount to a few $T$. This contribution
accounts for a larger $\beta$ leading to non-convexity. (The
peculiar dynamic surface tension behavior shown in 
Fig.~\ref{fig_plat_exp} is not
observed for {\em every} ionic surfactant. It has not been
observed, for example, in salt-free DTAB solutions
\cite{private}.)

\subsection{Adding Salt}

Finally, let us consider the effect of adding salt to an ionic
surfactant solution. For simplicity, and in accord with practical
conditions, it is assumed that the salt ions are much more mobile
than the surfactant and their concentration exceeds that of the
surfactant. In addition, we take the salt ions to be monovalent
and surface-inactive. Under these assumptions, the kinetics of the
salt ions can be neglected, reducing their role to the formation
of a thin electric double layer near the interface, which
maintains quasi-equilibrium with the adsorbed surface charge. The
double-layer potential is taken in the linear, Debye-H\"{u}ckel
regime \cite{AndelmanES}--\cite{DH},
$
  \Psi(x,t) = ({4\pi e\lambda}/{\varepsilon a^2}) \phi_0(t)
  \rme^{-x/\lambda}
$,
with a modified definition of the Debye-H\"{u}ckel screening length,
$\lambda \equiv (8\pi c_{\rm s} l)^{-1/2}$, $c_{\rm s}\gg c_{\rm b}$
being the salt concentration (the superscript `+' is omitted
hereafter from the surfactant symbols).

Substituting the double-layer potential in Eqs.~\ref{diffusion2} and 
\ref{dp1dt2}, the kinetic equations in the
bulk and sub-surface layer are obtained,
\begin{eqnarray}
 \frac{\partial\phi}{\partial t} &=& D \frac{\partial}{\partial x} \left(
         \frac{\partial\phi}{\partial x} - \frac{\phi_0{\rm e}^{-x/\lambda}}{2a^2\lambda^2 c_{\rm s}}
         \phi \right),
  \label{diffusion3} \\
  \frac{\partial\phi_1}{\partial t} &=& \frac{D}{a} \left( \left. \frac{\partial\phi}{\partial x}
         \right|_{x=a} - \frac{\phi_0}{2a^2\lambda^2 c_{\rm s}} \phi_1 \right) -
         \frac{\partial\phi_0}{\partial t},
  \label{dp1dt3}
\end{eqnarray}
whereas the kinetic equation at the interface itself remains the
same as Eq.~\ref{dp0dt2}.
Considering the electric potential as a small perturbation,
Eqs.~\ref{diffusion3} and \ref{dp1dt3} lead to the
asymptotic expression
\begin{eqnarray}
  \phi_1(t\rightarrow\infty)/\phi_{\rm b} &\simeq&
  1 - \phi_{0,\eq}/(2a^2\lambda c_{\rm s})
    - {(\tau_1/t)}^{1/2}
\nonumber \\
  \tau_1 &\equiv& \tau_1^{(0)} \left[ 1 - \frac{c_{\rm b}}{2c_{\rm s}} -
  \frac{\phi_{0,\eq}}{2a^2\lambda c_{\rm s}} \left( 1 - 
  \frac{3c_{\rm b}}{2c_{\rm s}} \right) \right]^2,
\label{asymdiffsalt}
\end{eqnarray}
where $\tau_1^{(0)}$ denotes the diffusion time scale in the
non-ionic case (Eq.~\ref{asymdiff}).
Due to surface charge, the equilibrium sub-surface
concentration is smaller than that of the bulk reservoir.
More important, though, is the correction to the diffusion time
scale introduced by the screened electrostatic interactions.
As expected, it decreases with increasing salt concentration.

Since the kinetic equation at the interface is identical to the one
in the absence of salt, so is the expression for the corresponding
time scale.
In the case of added salt, however,
the surface potential is much smaller than $T/e$, and the kinetic
time scale, $\tau_{\rm k}$, becomes only slightly larger than the
non-ionic one (Eq.~\ref{asymkin}).
Ionic surfactants with added salt are expected, therefore,
to behave much like
non-ionic surfactants, \ie undergo diffusion-limited adsorption
if no strong hindrance to adsorption exists.
The departure from the non-ionic behavior depends on salt
concentration and is described to first approximation by
Eq.~\ref{asymdiffsalt}.
The `footprint' of diffusion-limited adsorption, \ie a
$t^{-1/2}$ asymptotic time dependence, is observed in experiments,
as is demonstrated in Fig.~\ref{fig_salt_DLA}.
Consequently, the scheme described in previous sections for solving
the adsorption problem and calculating the dynamic surface tension
in the non-ionic case is applicable also for ionic surfactants
with added salt, and good fitting to experimental measurements
can be obtained \cite{Langevin2}.

\section{Summary}

We have reviewed a theoretical approach to the
fundamental problem of the adsorption kinetics of surfactants.
The formalism is more general than
previous ones as it yields the kinetics in the entire system,
both in the bulk solution and at the interface, relying on a
single functional and reducing the number of externally inserted
assumptions previously employed.

Common non-ionic surfactants, not hindered by high adsorption
barriers, are shown to undergo diffusion-limited
adsorption, in agreement with experiments.
In the non-ionic case our general formalism coincides with previous
ones and helps clarify the validity of their assumptions.
The adsorption process can be roughly divided into three
temporal stages.
At extremely early times the surface coverage and surface tension
change linearly with time because of interfacial
kinetics. This stage, however, is in most practical cases too short
to be observed experimentally (usually less than microseconds).
Due to this fast adsorption stage, the sub-surface layer becomes
nearly empty, which in turn drives a second,
diffusion-limited stage, where the surfactant
diffuses from the bulk with a $t^{1/2}$ time dependence.
The final relaxation towards equilibrium is usually
diffusion-limited, exhibiting an asymptotic $t^{-1/2}$ behavior.

In non-ionic surfactant mixtures, the initial adsorption stages
are dominated by the more mobile species. In cases where the less
mobile species is more surface-active, an intermediate stage is
predicted --- while one species undergoes desorption, the coverage
gradually becomes dominated by the other, energetically favorable
surfactant. The kinetic behavior of the mixture can be evaluated
based on equilibrium isotherms and single-surfactant data,
yielding good agreement with experiments. For surfact\-ant mixtures
exhibiting kinetically limited adsorption, we find a
`synergistic' effect, where the mixture kinetics may be
considerably different from that of the individual species. In
cases of high equilibrium surface coverage, a significant decrease
in adsorption rate is predicted due to coupling between the two
surfactants.

Strong electrostatic interactions in salt-free ionic surfactant
solutions are found to have a dramatic effect.
The adsorption becomes kinetically limited, which may lead to an
unusual time dependence, as observed in dynamic surface tension
measurements.
Such a scenario could not be accounted for by previous models.
Addition of salt to ionic surfactant solutions leads to screening
of the electrostatic interactions, and the adsorption becomes
similar to the non-ionic one, \ie diffusion-limited.
The departure from the non-ionic behavior as the salt concentration
is lowered has been described by a perturbative expansion.

A general method to calculate dynamic surface tension is obtained
from our formalism.
In the diffusion-limited case it coincides with previous results
which used the equilibrium equation of state.
In the kinetically limited case it produces different
expressions leading to novel conclusions.

Our kinetic model is restricted to simple relaxation processes,
where the free energy monotonously decreases with time.
In order to provide a quantitative treatment of more complicated
situations, such as the ones described for salt-free ionic solutions,
a more accurate theory is required, including, \eg a nucleation
mechanism.

Finally, as was demonstrated by the various cases treated in this
review, the approach presented here can be easily extended to
include additional components and interactions. 
This can be done by incorporating other terms in the 
excess free energy, Eq.~\ref{Dg_general}, and working out the 
kinetics in the same scheme as presented above.
Examples for interesting extensions are adsorption
from micellar solutions and the incorporation of
lateral diffusion.


\vspace{1cm}
\noindent
{\em Acknowledgments}

We thank E. Franses,
D. Langevin, R. Miller and C. Radke for
stimulating discussions and correspond\-ence.
Partial support from the Israel Science Foundation founded by
the Israel Academy of Sciences and Humanities -- Centers of
Excellence Program -- and the U.S.-Israel Binational Foundation
(B.S.F.) under grant no.~98-00429 is gratefully acknowledged.


\vspace{\baselineskip}
\begin{center}
{\bf Table 1.~~Comparison of the Predicted $\tau_\gamma$
to Experiment$^{\rm a}$}\\
\begin{tabular}{ccccccccc}
\hline
&&&&&&&& \\
  $ {\rm A} $
& $ {\rm B} $ & $ \phi_{0} / \bar{\phi_{0}} $ & $ \psi_{0} /
\bar{\psi_{0}} $ & $ \Delta\bar{\gamma}_{\rm A}\sqrt{
\bar{\tau}_{\rm A} } $ & $ \Delta\bar{\gamma}_{\rm B}\sqrt{
\bar{\tau}_{\rm B} } $ & $ \Delta\gamma\sqrt{\tau_\gamma}$ &
$\Delta\gamma\sqrt{\tau_\gamma}$ & error
\\
&&&&&& (th) & (exp) & \\
\hline
&&&&&&&& \\
X-405 & X-45 &
0.13 & 0.69& 0.6 & 62 & 29.5 & 32 & 8\% \\ &&&&&&&&
\\ X-405 & X-100 & 0.25 & 0.67 & 0.6 & 38 & 17.1 & 17 & 0.6\% \\
&&&&&&&& \\ X-405 & X-114 & 0.06 & 0.71& 0.6 & 14 & 7.1 & 6.8 &
4\% \\ &&&&&&&& \\ X-405 & X-165 & 0 & 1.4 & 0.6 & 4.4 & 8.6 & 6.5
& 33\%
\\ \hline
\end{tabular}
\parbox{5.75in}
{\setlength{\baselineskip}{18pt}
\vspace{\baselineskip}
$^{\rm a}$The materials used were sequences of Triton~X
mixtures \cite{Fainerman2}.
The single-surfactant values, $\bar{\phi}_0$,
$\bar{\psi}_0$, $\Delta\bar{\gamma}_{\rm A}\sqrt{\bar{\tau}_{\rm
A}}$, $\Delta\bar{\gamma}_{\rm B}\sqrt{\bar{\tau}_{\rm B}} $, and
equilibrium coverages for the mixture, $\phi_{0,{\rm eq}}$ and
$\psi_{0,{\rm eq}}$, are taken from the same reference. The values
for $\Delta\bar{\gamma}\sqrt{\tau_{\gamma}}$ (given in units of
dyn s$^{1/2}$/cm) are obtained experimentally from the asymptotic
slope of $\gamma$ vs. $t^{-1/2}$ curves (see
Eq.~\ref{various_tau}). The predicted values for
$\Delta\bar{\gamma}\sqrt{\tau_{\gamma}}$ of the mixture and the
corresponding experimental results are given in the columns
indicated by `th' and `exp', respectively. The last column shows
the respective error between theory and experiment.}
\end{center}
\vspace{1\baselineskip}

\newpage
\section*{Figure Captions}

\begin{figure}[p]
\centerline{
  \resizebox{.57\textwidth}{!}
  {\includegraphics{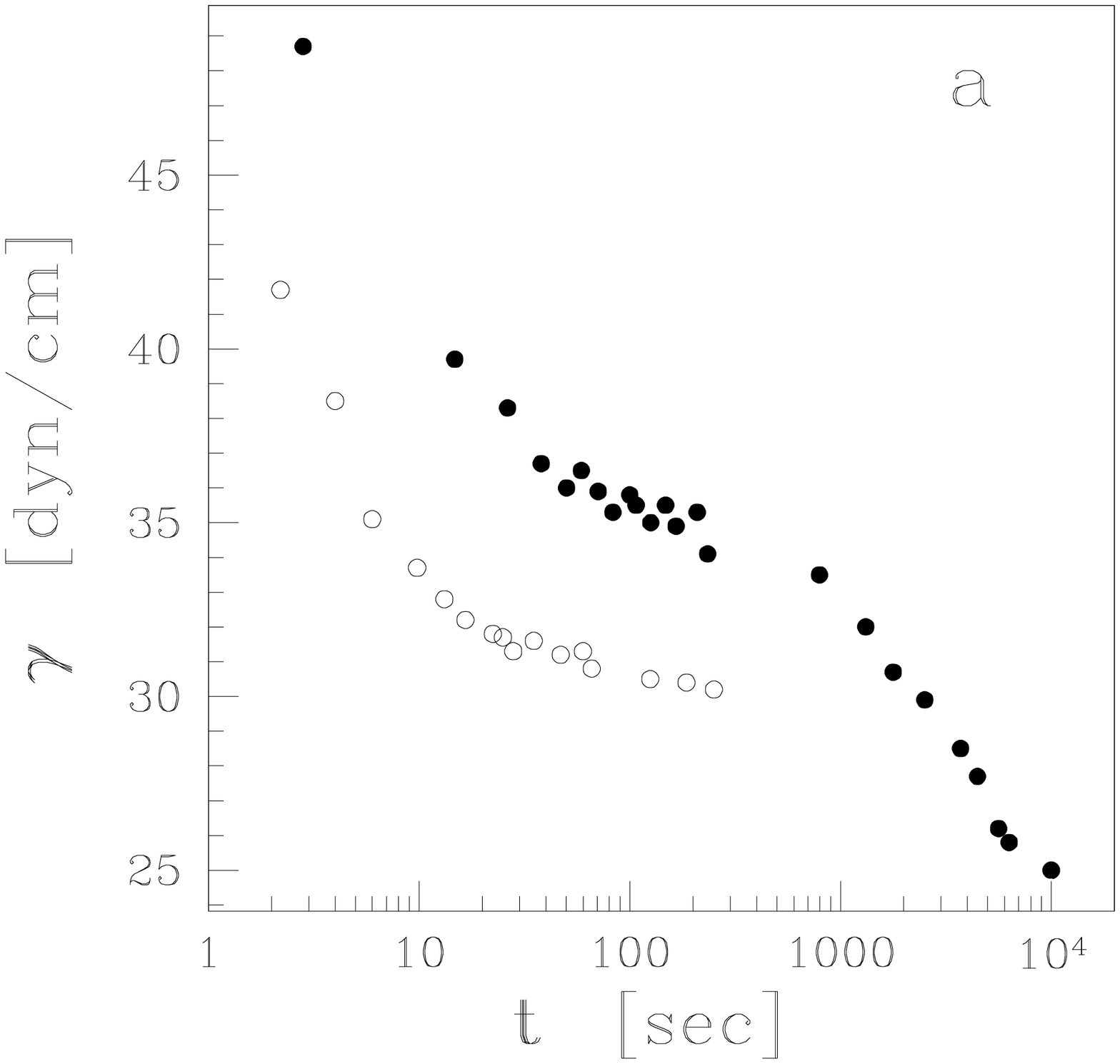}}
  \resizebox{.57\textwidth}{!}
  {\includegraphics{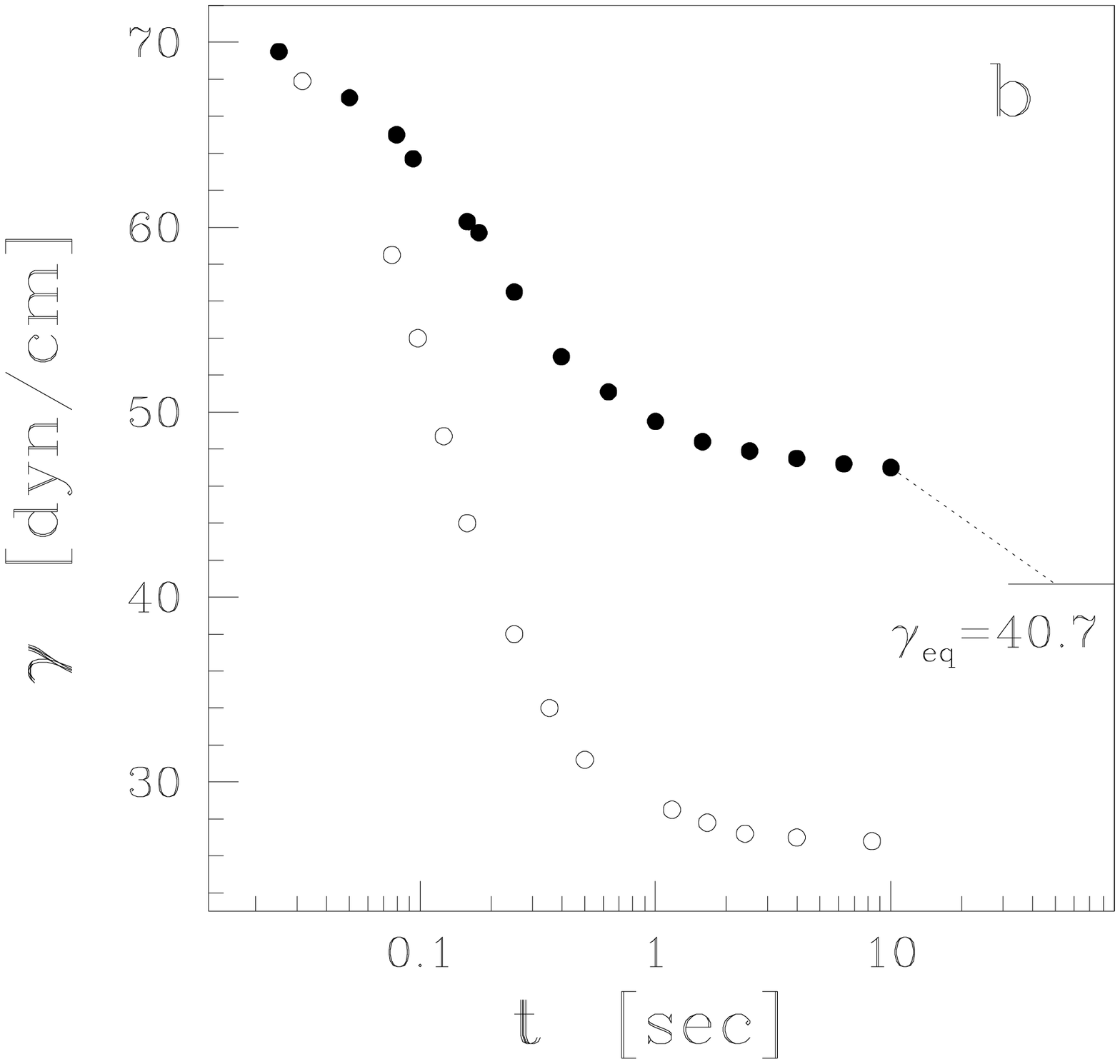}}}
\caption[]{}
\label{fig_plat_exp}
\end{figure}

\begin{figure}[p]
\centerline{ \resizebox{.57\textwidth}{!}
{\includegraphics{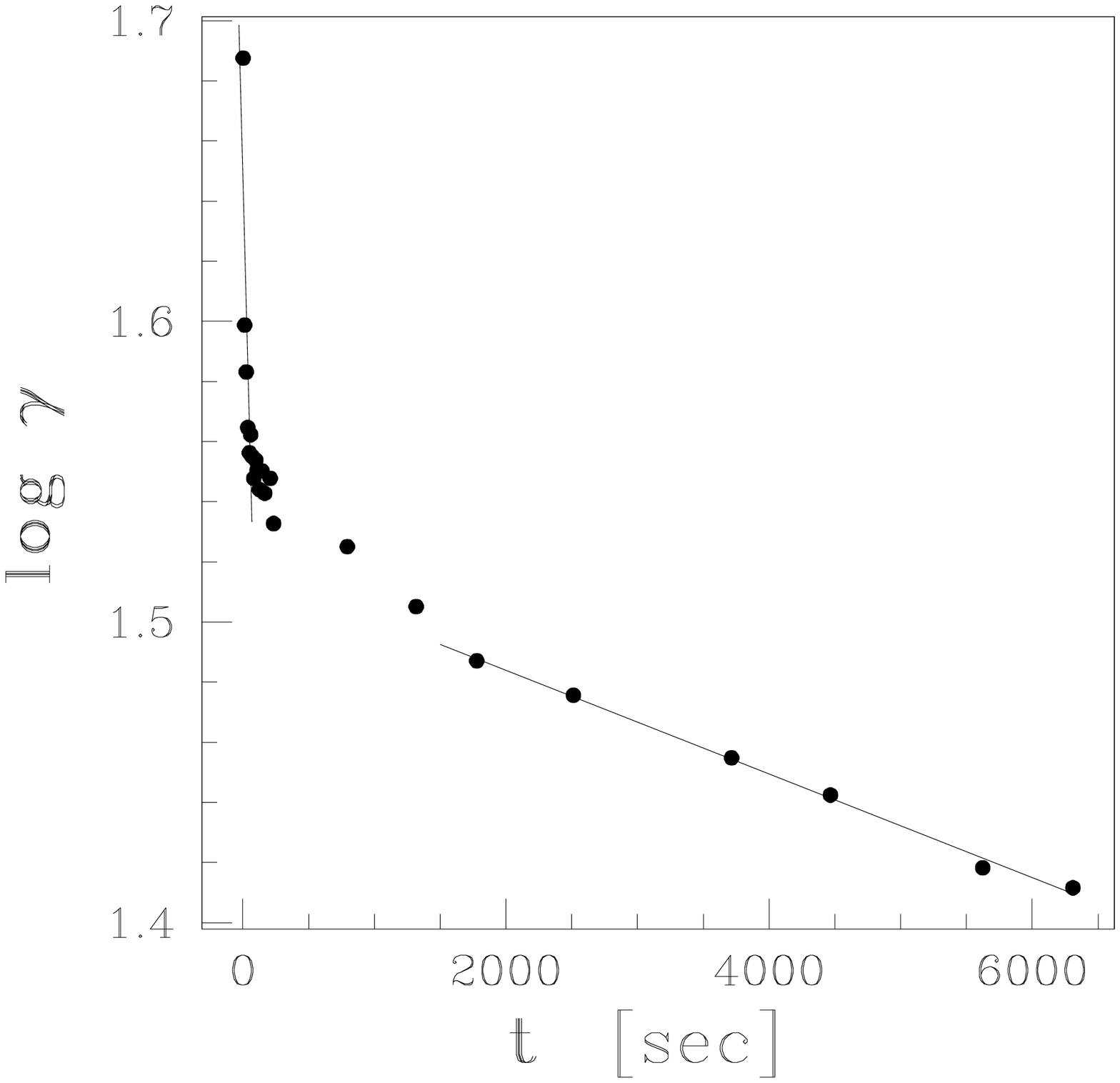}}}
\caption[]{}
\label{fig_exponent}
\end{figure}

\begin{figure}[p]
\centerline{
\resizebox{.57\textwidth}{!}
{\includegraphics{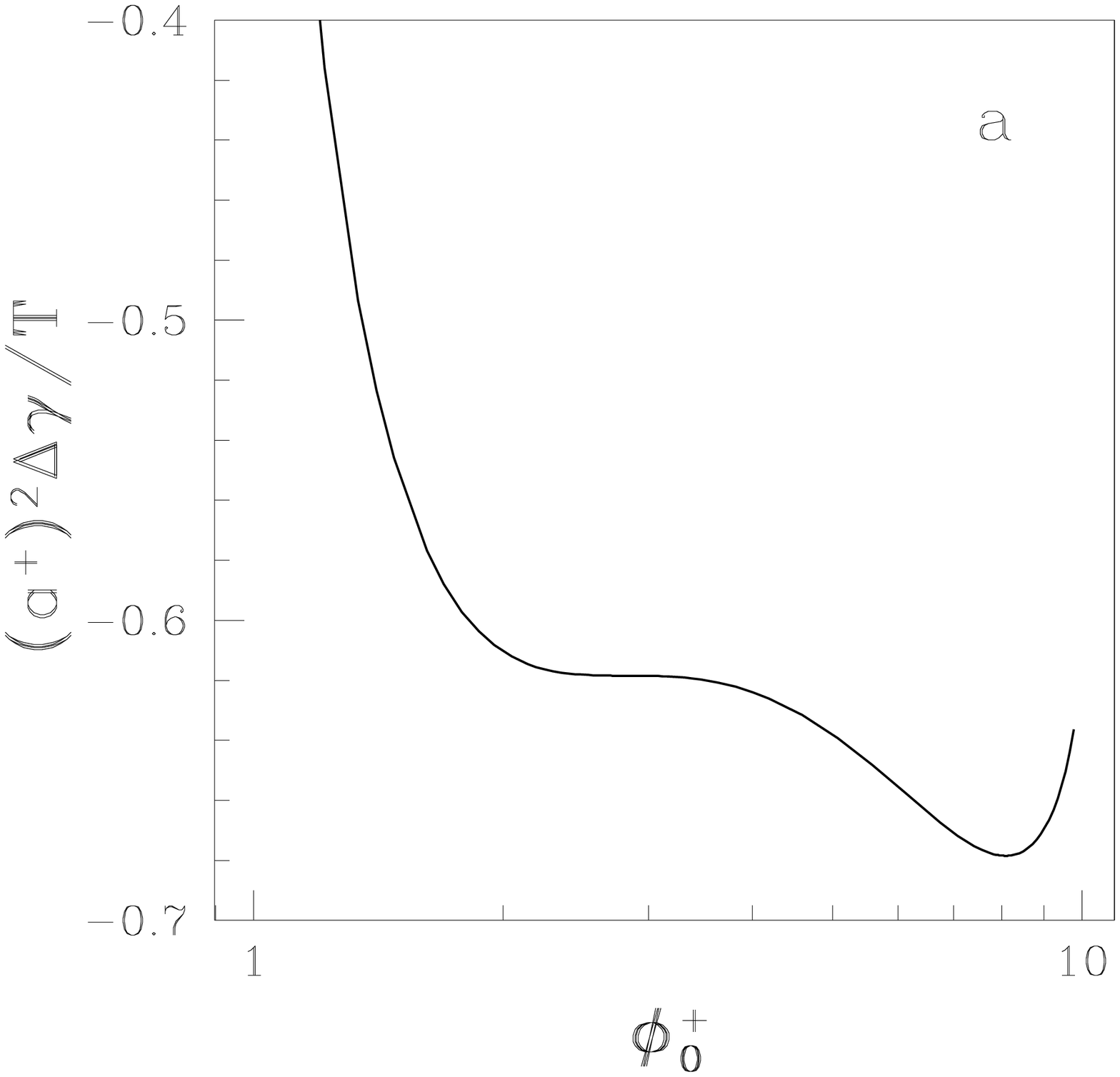}}
\resizebox{.57\textwidth}{!}
{\includegraphics{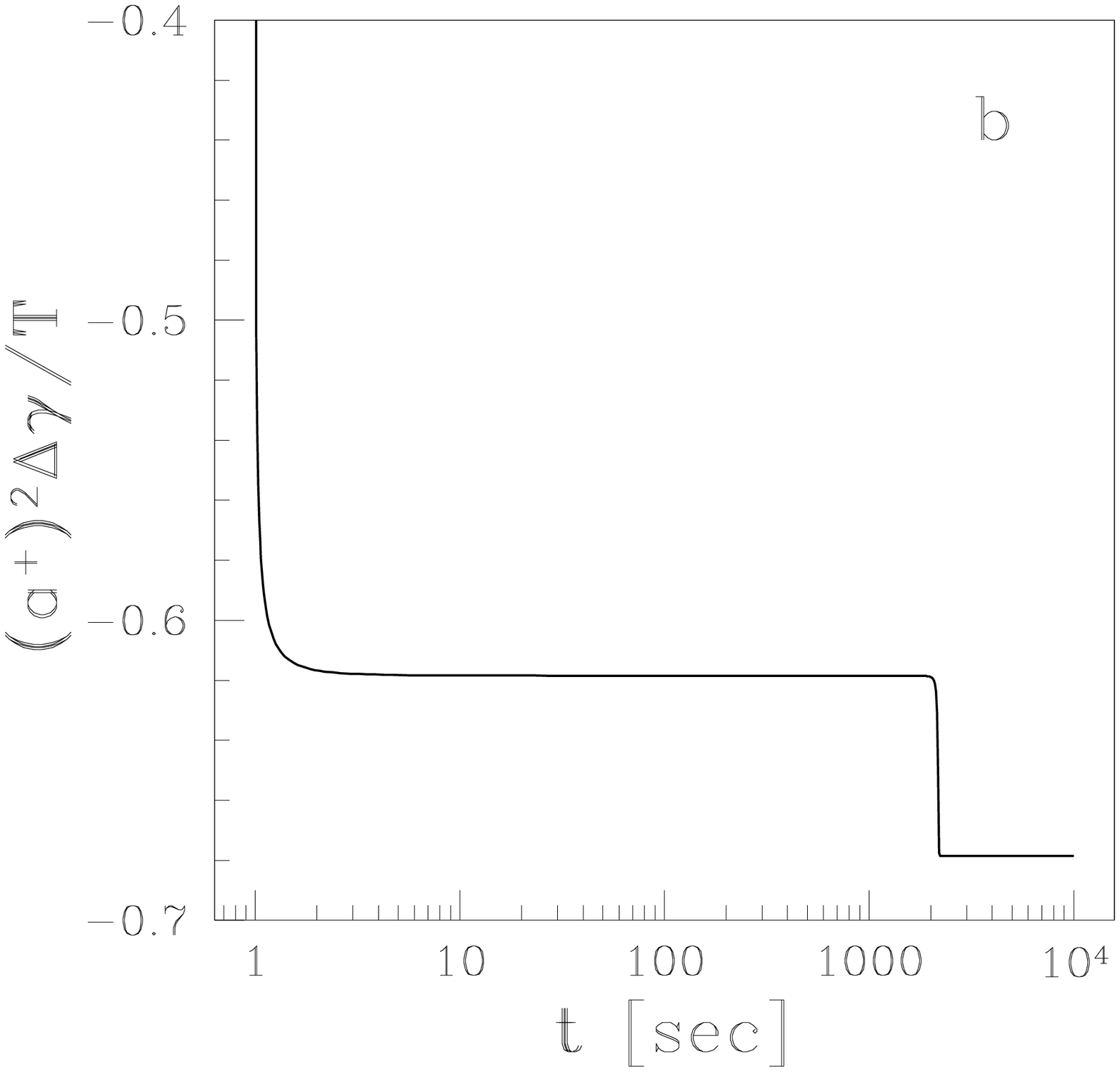}}}
\caption[]{}
\label{fig_plat_theo}
\end{figure}

\begin{figure}[p]
\centerline{ \resizebox{.57\textwidth}{!}
{\includegraphics{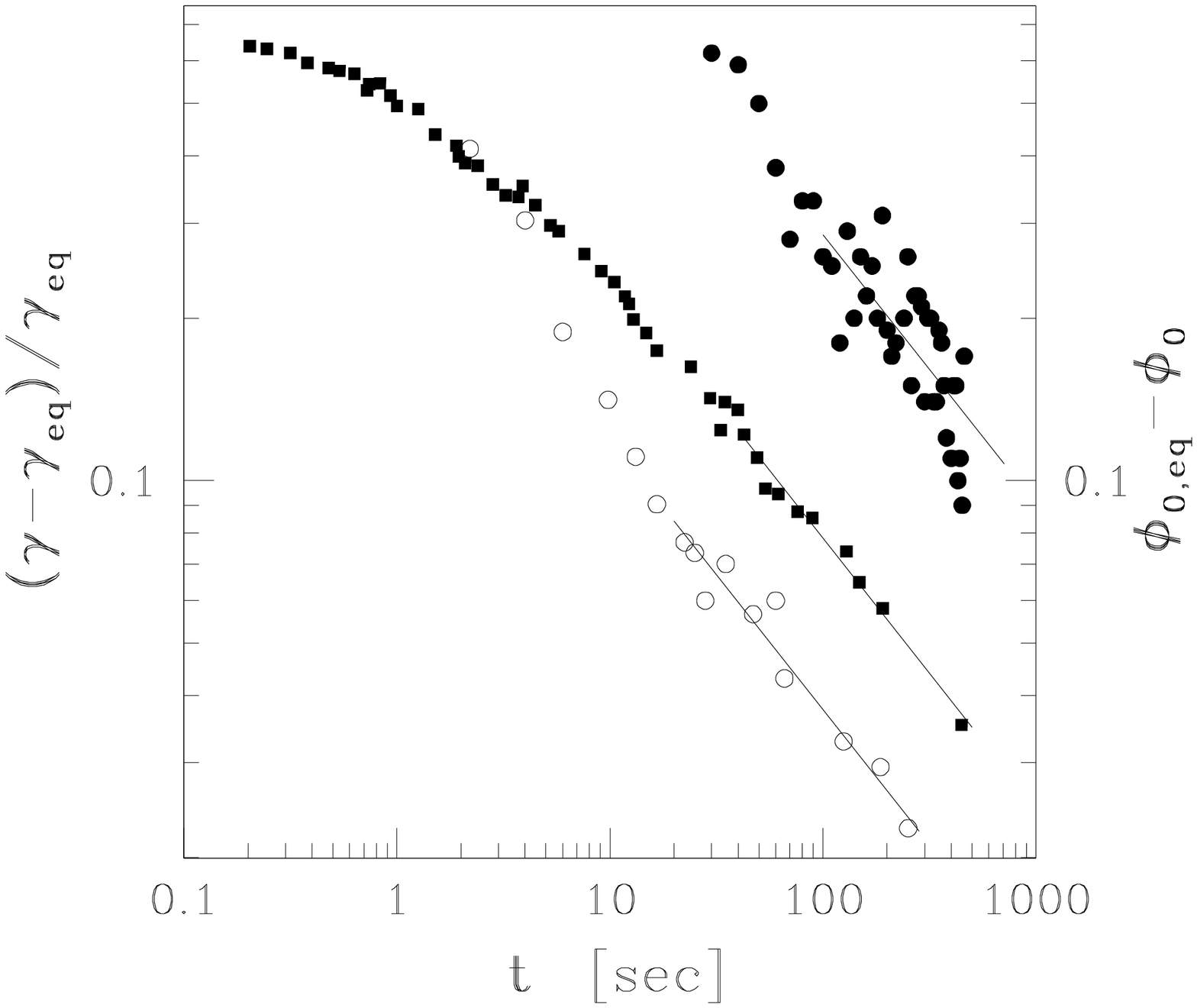}}}
\caption{}
\label{fig_salt_DLA}
\end{figure}

\begin{itemize}

\item[Fig. 1]
Schematic view of the system. A sharp,
flat interface separates a dilute surfactant solution from an air
or oil phase.

\item[Fig. 2]
Typical dynamic surface tension curve of a non-ionic
  surfactant solution. (Adapted from Ref.~\cite{Lin2}).
   The solution contains $1.586\times 10^{-4}$M 1-decanol.
   The solid line is a theoretical fit using the following
   parameters:
   $a=4.86$ \AA, $\alpha=11.6 T$, $\beta=3.90 T$
   (all three parameters
   were fitted
   from independent equilibrium measurements), and
   $D=6.75\times 10^{-6}$ cm$^2$/sec.

\item[Fig. 3]
Diffusion-limited adsorption
exhibited by non-ionic surfactants.
Four examples for dynamic surface tension measurements are shown.
Open circles --- decyl alcohol,
$9.49\times 10^{-5}$M. (Adapted from Ref.~\cite{AddHutch}.)
Squares --- Triton X-100, $2.32\times 10^{-5}$M.
(Adapted from Ref.~\cite{Lin1}.)
Triangles --- C$_{12}$EO$_8$, $6\times 10^{-5}$M.
(Adapted from Ref.~\cite{Hua2}.)
Solid circles --- C$_{10}$PY, $4.35\times 10^{-4}$M.
(Adapted from Ref.~\cite{Hua2}.)
The asymptotic $t^{-1/2}$ dependence shown by the solid fitting
lines is a `footprint' of diffusion-limited adsorption.

\item[Fig. 4]
Dependence of surface tension on surface
 coverage in
 diffusion-limited adsorption (Eq.~\ref{eqstate}).
 The values taken for the parameters match the example in
 Figure~1.

\item[Fig. 5]
(a) Surface coverage in a mixture of
interacting surfactants. The dotted, dashed and solid lines are
the surface coverages of surfactants A ($\phi_0$), B ($\psi_0$),
and the total coverage ($\phi_0+\psi_0$), respectively. The
assigned parameters are: $\phi_{\rm b} = 10^{-4}$, $\psi_{\rm b} =
2\times 10^{-4}$, $\alpha_{\rm A} = 10T $, $\alpha_{\rm B} = 9T $,
$\beta_{\rm A} = \beta_{\rm B} = 3T $, $\chi = T$,
$D_{\rm A}^{1/2} / a = 300 \; {\rm s}^{-1/2} $, and
$D_{\rm B}^{1/2}  / a =
900 \; {\rm s}^{-1/2} $. This implies that surfactant A diffuses
more slowly but is more surface active.
(b) Dynamic surface tension of the same system.

\item[Fig. 6]
(a) Dynamic interfacial tension between SDS aqueous solutions and
  dodecane.
Filled circles --- $3.5\times 10^{-4}$M SDS without salt;
open circles --- $4.86\times 10^{-5}$M SDS with 0.1M NaCl.
(Adapted from Ref.~\cite{Langevin2}.)
  (b) Dynamic surface tension between $5.84\times 10^{-4}$M
  DESS solution and air.
Filled circles --- without salt;
open circles --- with 0.1M NaCl.
(Adapted from Ref.~\cite{HuaDESS};
the authors did not provide
details of the relaxation towards final equilibrium in the
salt-free case.)

\item[Fig. 7]
Dynamic surface tension of the salt-free SDS solution
  of Fig.~\ref{fig_plat_exp}a, redrawn on a semi-log plot.
  Two exponential relaxations are observed, indicating a kinetically
  limited process.

\item[Fig. 8]
(a) Dependence of surface tension on surface coverage
  in kinetically limited adsorption (Eq.~\ref{Dgkinetic}).
  The values taken for the parameters are:
  $a^+=17$ \AA, $\phi^+_{\rm b}=6\times 10^{-5}$, $\alpha=14.78T$ and
  $\beta=8.5T$. The values were selected to yield a non-convex,
  yet decreasing curve (see text).
  (b) The corresponding dynamic surface tension, calculated using
  Eqs.~\ref{dp0dt2_kin} and \ref{Dgkinetic} with the value
  $D^+_0=6\times 10^{-6}$ cm$^2$/sec.

\item[Fig. 9]
   Diffusion-limited adsorption exhibited by ionic
   surfactants with added salt.
   Open circles and left ordinate ---
   dynamic interfacial tension between  dodecane and an aqueous solution 
   of $4.86\times 10^{-5}$M SDS with 0.1M NaCl.
   (Adapted from Ref.~\cite{Langevin2}.)
   Squares and left ordinate ---
   dynamic surface tension of an aqueous solution of
   $2.0\times 10^{-4}$M
   SDS with 0.5M NaCl.
   (Adapted from Ref.~\cite{Fainerman3}.)
   Filled circles and right ordinate ---
   surface coverage deduced from second harmonic generation
   measurements on a saturated aqueous solution of SDNS with
   2\% NaCl. (Adapted from Ref.~\cite{SHG}.)
   The asymptotic $t^{-1/2}$ dependence shown by the solid
   fitting lines is a `footprint' of diffusion-limited
   adsorption.

\end{itemize}

\newpage

%

\end{document}